\documentclass[aps,prd,preprint,superscriptaddress,showkeys]{revtex4-2}
\usepackage{lastpage}
\usepackage{colortbl}
\usepackage{totcount}
\usepackage{changepage}
\usepackage{todonotes}
\usepackage{amsmath}
\usetikzlibrary{decorations.pathmorphing}
\usetikzlibrary{decorations.markings}
\tikzset{
    photon/.style={decorate, decoration={snake}},
    electron/.style={postaction={decorate},
        decoration={markings,mark=at position .55 with {\arrow[draw=blue]{>}}}},
    gluon/.style={decorate,decoration={coil,amplitude=4pt, segment length=5pt}} 
}
\newcommand{\condensat}[2]{\filldraw[white] ({#1-0.1},{#2-0.25}) rectangle ++(0.2,0.5);\draw ({#1-0.1},{#2-0.25}) -- ++(0,0.5);\draw ({#1+0.1},{#2-0.25}) -- ++(0,0.5);}
\newcommand{\fletxa}[3]{\draw[->] (#1,#2) -- ++({0.001*cos(#3)},{0.001*sin(#3)});}
\newcommand{\rodona}[2]{\filldraw[white] (#1,#2) circle (3pt); \draw (#1,#2) circle (3pt);}
\newcommand{\quadrat}[2]{\filldraw[white] ({#1-0.1},{#2-0.1}) rectangle ++(0.2,0.2); \draw ({#1-0.1},{#2-0.1}) rectangle ++(0.2,0.2);}

\begin{document}
\title{AN EFFECTIVE FIELD THEORY STUDY OF MEDIUM HEAVY QUARK EVOLUTION}
\author{Miguel \'{A}ngel Escobedo}
\email[]{miguel.a.escobedo@fqa.ub.edu}
\affiliation{Departament de F\'{i}sica Qu\`{a}ntica i Astrof\'{i}sica and Institut de Ci\`{e}ncies del Cosmos, Universitat de Barcelona, Martí i Franqu\`{e}s 1, 08028 Barcelona, Catalonia, Spain.} 
\begin{abstract}
    The evolution of hard probes in a medium is a complex multiscale problem that significantly benefits from the use of Effective Field Theories (EFTs). Within the EFT framework, we aim to define a series of EFTs in a way that addresses each energy scale individually in separate steps. However, studying hard probes in a medium presents challenges. This is because an EFT is typically constructed by formulating the most general Lagrangian compatible with the problem's symmetries. Nevertheless, medium effects may not always be encoded adequately in an effective action. In this paper, we construct an EFT that is valid for studying the evolution of a heavy quark in a QCD plasma containing few other heavy quarks, where degrees of freedom with an energy of the order of the temperature scale are integrated out. Through this example, we explicitly demonstrate how to handle the doubling of degrees that arise in non-equilibrium field theory. As a result, we derive a Fokker-Planck equation using only symmetry and power counting arguments. The methods introduced in this paper will pave the way for future developments in the study of quarkonium suppression.
\end{abstract}
\date{\today}
\maketitle
\section{Introduction}
The quark-gluon plasma is a state of matter that forms at high temperatures and densities, in which quarks and gluons are not confined within hadrons. This state of matter can be created on Earth in experiments using ultrarelativistic heavy-ion collisions. However, the quark-gluon plasma exists only for a very short time during these collisions, so we must study the particles produced during this brief period to learn about the properties of the plasma. One approach is to study ``hard probes," which are observables that are both significantly affected by the medium and can be measured in the challenging environment of a heavy-ion collision. Examples of hard probes include heavy quarks, heavy quarkonium, and jets.

Studying heavy particles and jets in a medium requires dealing with largely separated energy scales. Particles with energies of the order of the heavy quark mass or a hard parton energy are rare in the medium and can be accurately described using perturbative QCD. However, particles with energies of the order of the temperature are sensitive to the medium and cannot always be described using perturbation theory. Therefore, it is interesting to separate the contribution from these two kind of particles to encode non-perturbative effects in parameters or functions that can be computed using non-perturbative methods like lattice QCD. It is also important to be careful when describing systems with largely separated energy scales. On one hand, the appearance of largely separated energy scales can lead to a breaking of naive perturbation theory, meaning that the size of a contribution cannot be directly related to its number of loops. On the other hand, non-perturbative studies using lattice QCD are also challenging because a very large lattice is required to accommodate all the energy scales. 

These problems can be solved by using EFTs. An EFT is a quantum field theory that gives the same results as another more general theory at low energies. They are constructed in the following way \cite{Weinberg:1978kz}:
\begin{itemize}
\item Identify the relevant degrees of freedom and the symmetries of the problem.
\item Write the more general Lagrangian that respects the symmetries of the problem using the relevant degrees of freedom. 
\item An EFT must be equipped with a power counting. This means that there is a simple rule to predict how large is the contribution of each term in the Lagrangian for a given observable.
\item The unknown parameters in the effective Lagrangian are called Wilson coefficients. They are fixed by imposing that the EFT gives the same results as the full theory at low energies. This procedure is called matching.
\end{itemize}
Note that an EFT Lagrangian has an infinite number of terms. However, the theory still has predictive power thanks to the power counting.

The use of EFTs can also allow relating physical observables with quantities computable on the lattice in a more direct way. A good example is the description of heavy quarkonium. Heavy quarkonium is a bound state made of a heavy quark and a heavy antiquark. Quarkonium is a non-relativistic system in which well separated energy scales appear. They are the mass, $M$, the inverse of the typical radius, $1/r\sim Mv$ with $v\ll 1$, and the binding energy $E\sim Mv^2$. Non-relativistic QCD (NRQCD) \cite{Caswell:1985ui,Bodwin:1994jh} is an EFT valid for energies smaller than $M$. Since $M\gg\Lambda_{QCD}$ we can reliably match QCD to NRQCD using perturbation theory. We can also use Potential NRQCD (pNRQCD) \cite{Pineda:1997bj,Brambilla:1999xf,Brambilla:2004jw}, an EFT valid for energies much smaller than $1/r$. Since this scale is not always perturbative, there are cases in which the matching between NRQCD and pNRQCD cannot be done in perturbation theory. However, since each term in the pNRQCD Lagrangian has a scaling with $1/M$ that is easy to determine, we can perform the matching between NRQCD and pNRQCD in the limit in which the mass of the heavy quarks is infinite (the static limit). Once the matching is done, we can use these Wilson coefficients to compute the properties of quarkonium states with finite heavy quark masses. This is a rigorous way to show that the spectrum of quarkonium can be determined by solving a Schr\"{o}dinger equation in which the potential is computed on the lattice using static quarks. 

EFTs have been applied to the study of hard probes in heavy ion collisions. In the case of heavy quarkonium, medium modified versions of NRQCD and pNRQCD has been used to compute the thermal corrections to the mass and the medium induced decay width \cite{Escobedo:2008sy,Brambilla:2008cx,Brambilla:2010vq}. pNRQCD has also been used to study the evolution of the population of bound states inside of a medium in the cases in which $1/r\gg T$ \cite{Brambilla:2016wgg,Brambilla:2017zei,Yao:2017fuc,Yao:2018nmy}. Regarding jets, there are recently developed EFTs for the study of jet broadening in \cite{Vaidya:2020cyi} and jet sub-structure in heavy-ion collisions \cite{Vaidya:2020lih}. 

\begin{figure}
    \begin{center}
    \begin{tikzpicture}
        \draw[->] (-4,0) -- (4,0) node[above] {$\textrm{Re}\, t$};
        \draw[->] (0,-4) -- (0,4) node[right] {$\textrm{Im}\, t$};
        \draw[very thick] (-3.8,0) -- (3.8,0) -- (3.8,-0.2) -- (-3.8,-0.2);
    \end{tikzpicture}
\end{center}
    \caption{Representation of the Schwinger-Keldysh contour. The thicker line represents the complex time contour along which the path integral is defined in non-equilibrium field theory. It goes from $-\infty$ to $\infty$ and then goes back to $-\infty$ but decreasing the time's imaginary part by an infinitely small amount.}
    \label{fig:sk}
\end{figure}
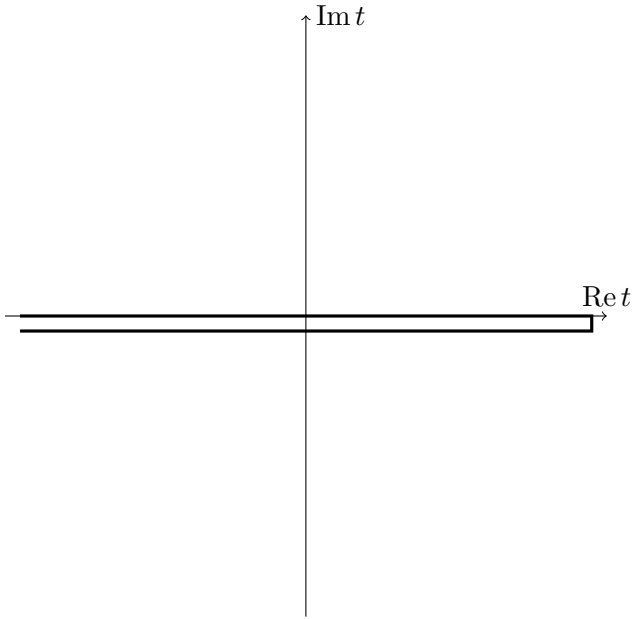

However, the application of the EFT framework to the study of hard probes in a medium presents a conceptual challenge. The procedure that we have outlined before needs to be modified when the medium affects the matching between the full theory and the EFT. At finite temperature, computations of non-static properties are done using the so-called real-time formalism \cite{Bellac:2011kqa}. In this formalism, there is a doubling of degrees of freedom. This means that the standard path integral has to be substituted by a path integral in which the time integration follows the complex Schwinger-Keldysh contour, see fig. \ref{fig:sk}. In practice, instead of considering explicitly the complex-time integration, we name fields with a time argument on the upper (lower) branch of the contour fields of type 1(2). As an example, let us consider a QFT describing the evolution of a field $\phi$. Then the action can be written as
\begin{equation}
    S=\int_C\,dt L[\phi(t)]=\int_{-\infty}^\infty\,dt L[\phi_1(t)]-\int_{-\infty}^\infty\,dt L[\phi_2(t)]\,,
    \label{eq:Ssk}
\end{equation}
where the subindex $C$ means that the integration is done along the Schwinger-Keldysh contour. In this way, due to the properties of the path integral, correlators in which only fields of type 1 (2) appear are chronologically (anti-chronologically) ordered. However, we might be interested in evaluating correlators that are neither chronologically nor anti-chronologically ordered. For example, if we want to have access to the distribution function we would need to compute $\langle\phi(t)\phi(0)\rangle$, where the ordering of the operators is as written. Note also that, although in eq. (\ref{eq:Ssk}) there are no terms containing at the same time both type of fields, the propagator of the field $\phi_i$ is not diagonal in the $1, 2$ indices \cite{Bellac:2011kqa}. 

What we have explained until now regarding the Schwinger-Keldysh contour (for the specific case of a QFT describing the evolution of a field $\phi$) is standard textbook material. However, it was important to remind it in order to highlight non-trivial features that appear when we apply the EFT framework to a non-equilibrium or thermal field theory. When we integrate out degrees of freedom, the matching between the full theory and the EFT might induce terms including the two types of fields. In other words, the EFT might not show the structure that appears in eq. (\ref{eq:Ssk}). Physically, we might interpret this in the following way. The EFT describes the evolution of an open quantum system interacting with an environment (the medium degrees of freedom that have been integrated out). It is a well-known result in quantum mechanics that the evolution of an open quantum system can not be encoded in a Hamiltonian \cite{Breuer:2007juk}. In other words, in general it does not exist an operator such that the time-evolution of the density matrix can be written as a von-Neumann equation. Instead, the evolution of the density matrix of a open quantum system follows a more general kind of equation. The same is true if we apply the path integral formalism. The effect of high energy fields on low energy ones can not be encoded in an effective action when we are dealing with an open quantum system. We need a more general object called influence functional instead \cite{Feynman:1963fq}. For the example of a $\phi$ field
\begin{equation}
    \langle\mathcal{P} \phi_i(t)\phi(t')\rangle=\int\mathcal{D}\phi_1\mathcal{D}\phi_2\mathcal{F}[\phi_1,\phi_2]\phi_i(t)\phi_j(t')\,,
\end{equation}
where $\mathcal{P}$ means ordering along the Schwinger-Keldysh contour and $\mathcal{F}$ is the influence functional. Only in the case of an isolated system it happens that $\mathcal{F}=\mathrm{e}^{S[\phi_1]-S[\phi_2]}$, where $S$ is the effective action.

Taking this into account is a substantial modification of the second point of the procedure we outlined to construct and EFT. Up to now, this difficulty has been skipped in the study of hard probes using two different strategies. The first one is to focus on computing the binding energy and the decay width of non-relativisty systems. In this case, the effects of the doubling of degrees of freedom are neutralized in the approximation in which the heavy particles are dilute \cite{Escobedo:2008sy,Brambilla:2008cx}. The second strategy is to study cases in which the temperature is smaller than the energies that are integrated out to define the EFT \cite{Brambilla:2016wgg,Brambilla:2017zei}.

The construction of EFTs that do not consist in an effective action has been investigated in the context of the development of an EFT for hydrodynamics \cite{Baidya:2017eho,Haehl:2016pec,Crossley:2015evo,Jensen:2017kzi}. However, to our knowledge, these developments have not been applied to the study of hard probes in a QCD plasma. In this manuscript, we aim to fill this gap by studying a simple example. A heavy quark close to thermalization in a medium with few other heavy quarks. In this problem, we can identify three widely separated energy scales.
\begin{itemize}
\item The mass of the heavy quark $M$.
\item The spatial momentum of the heavy quark $p$, which is of order of $\sqrt{TM}$, being $T$ the temperature.
\item The temperature $T$ and other energy scales that the medium might induced.
\end{itemize}
It is well-known that this situation can be studied using a Fokker-Planck equation or the physically equivalent Langevin equation \cite{Svetitsky:1987gq,Torres-Rincon:2012sda,Moore:2004tg,vanHees:2005wb,Rapp:2008qc,Akamatsu:2008ge,Young:2008he}. In this manuscript, we will give an EFT perspective on the problem. Following the EFT philosophy, we will deal with each energy scale in a separate way. We will follow these steps:
\begin{enumerate}
\item As a starting point, we can use NRQCD. This automatically encodes the effects of degrees of freedom with energy of order $M$.
\item It is straight-forward to define an EFT for energies smaller than $\sqrt{MT}$. This EFT is a modification of NRQCD using the momentum-label technique common in Soft-Collinear Effective Theory (SCET) \cite{Bauer:2000yr,Bauer:2001yt} and On-Shell Effective Theory (OSEFT) \cite{Manuel:2014dza,Manuel:2016wqs}. An equivalent EFT, with the same degrees of freedom, symmetries and power counting, was previously introduced to study a completely different problem. This is NRQCD for semi-hard fields ($\textrm{NRQCD}_{sh}$) \cite{Brambilla:2003mu}. This EFT was originally introduced to study modes with energy of the order $\sqrt{M\Lambda_{QCD}}$ in the study of heavy quarkonium at $T=0$. In our case, the role of $\Lambda_{QCD}$ is played by $T$. However, symmetries and power counting arguments remain the same. 
\item We define an EFT for energies below $T$. In this EFT there are terms that mix fields of type $1$ and type $2$. These are dealt with following the approach of \cite{Baidya:2017eho} but supplementing it with the properties of the dilute expansion. We call this EFT Langevin Effective Theory (LET) due to its connection with the Langevin equation.
\item We perform the matching between $\textrm{NRQCD}_{sh}$ and LET in the one-gluon exchange approximation as an illustration. This allows to encode the influence of degrees of freedom with energy of order $T$.
\item We use LET to compute the evolution of the density of heavy quarks, obtaining a Fokker-Planck equation for the density that coincides with the evolution resulting from a Langevin equation.
\end{enumerate}
This procedure could seem extremely complex to obtain results that are already well-known. However, our aim is to develop techniques that could be useful for more interesting cases. In particular, we have in mind the study of the evolution of the reduced density matrix of quarkonium in the regime $T\sim\frac{1}{r}$. The problem that we study in this manuscript is simpler than quarkonium in the regime $T\sim\frac{1}{r}$. However, in both cases we are dealing with dilute non-relativistic particles. One important observation that the framework developed here makes manifest is that the physics at the scale $T$ can be studied in the static limit. In other words, the matching between $\textrm{NRQCD}_{sh}$ and LET can be done in the static limit, and we can use those Wilson coefficients to study the finite $M$ case. A similar thing happens regarding the matching between NRQCD and pNRQCD at $T=0$. These two results combined strongly support the idea that the matching between NRQCD and pNRQCD at finite temperature can be also done in the static limit. We note that the evolution of a static heavy quark-antiquark pair in terms of gauge-invariant expectation values has already been discussed in \cite{Escobedo:2020tuc}. 

As a complement to the previous study, we will also discuss the case in which $p\sim T$. In this case, we integrate out the scales $p$ and $T$ at the same time. In other words, we go directly from NRQCD to LET. We observe that the structure of LET in this case is still the same as in the case $p\gg T$. However, the Wilson coefficients are different. We observe that there is a smooth transition between the cases $p\sim T$ and $p\gg T$, as it should. Physically, if we start with a heavy quark at rest, it will start to gain momentum due to broadening. As the momentum becomes larger, two complementary things happen. On one hand, the Fokker-Planck equation becomes accurate. On the other hand, the drag force becomes a leading order effect.

The present work has some similarities with \cite{Bu:2021jlp}, however the perspective is quite different. In \cite{Bu:2021jlp} the AdS/CFT correspondence in Supersymmetric Yang-Mills is taken as starting point, while in this work we tried to be as agnostic as possible about the properties of the medium.

The manuscript is organized as follows. In section \ref{sec:NRQCDl}, we will discuss $\textrm{NRQCD}_{sh}$. Next, in section \ref{sec:let} , we will present LET. Section \ref{sec:finlet} discusses the computation of the evolution of the distribution of heavy quarks within LET. In section \ref{sec:matching}, we discuss the matching between  $\textrm{NRQCD}_{sh}$ and LET. In the section \ref{sec:psimT}, we consider the case of a heavy quark with momentum $p\sim T$. Finally, in section \ref{sec:concl}, we give our conclusions.
\section{$\textrm{NRQCD}_{sh}$}
\label{sec:NRQCDl}
Our starting point is the NRQCD Lagrangian \cite{Caswell:1985ui,Bodwin:1994jh}. To fix the notation, we write its heavy quark sector
\begin{equation}
\mathcal{L}_\psi=\psi^\dagger(x)\{iD_0+\frac{\mathbf{D}^2}{2M}+c_4\frac{\mathbf{D}^4}{8M^3}+c_Fg\frac{\boldsymbol{\sigma}\cdot\mathbf{B}}{2M}\}\psi(x)+\cdots
\label{eq:NRQCD}
\end{equation}
Here $D_\mu=\partial_\mu-igA_\mu$, $\sigma$ is a Pauli matrix and $E$ and $B$ are the chromoelectric and chromomagnetic fields. We use the following power counting. The spatial momentum of the heavy quark scales like $\sqrt{MT}$ plus a possible residual momentum of order $T$. The gauge field is only sensitive to the scale $T$. In eq. (\ref{eq:NRQCD}) we have only written terms of order $\psi^\dagger\psi\frac{T^2}{M}$ or lower. These are the terms that we need in order to compute $\frac{1}{f}\frac{df}{dt}$ to order $\frac{T^2}{M}$ where $f(p,\mathbf{R})$ is the distribution function of heavy quarks. However, only the first two terms in the Lagrangian will end up contributing. We define $f$ in terms of the < propagator of heavy quarks. More details will be given later.

Using NRQCD to study this problem is not completely optimal. As we mentioned before, the spatial momentum of heavy quarks is of order $\sqrt{MT}$ while each interaction with the medium changes the momentum of the heavy quark by an amount of order $T$. This implies that the term $\psi^\dagger\frac{\mathbf{D}^2}{2M}\psi$ hides contributions of different sizes. Then the power counting is not completely clear. Each time that we apply this term it is not obvious whether we will get a contribution of order $T$, $T\sqrt{\frac{T}{M}}$ or $\frac{T^2}{M}$. At best, we can put an upper bound on the size of the contribution. To improve the situation we can introduce momentum-label fields as it is done in SCET and OSEFT. Let us divide the spatial momentum of heavy quarks in two pieces
\begin{equation}
\mathbf{p}+\mathbf{k}\,,
\end{equation}
where $\mathbf{p}$ is of order $\sqrt{MT}$ and $\mathbf{k}$ is a residual momentum of order $T$. To improve this we can perform the following transformation
\begin{equation}
\psi(t,\mathbf{x})=\sum_{\mathbf{p}}\mathrm{e}^{i\mathbf{p}\mathbf{x}}\xi_\mathbf{p}(t,\mathbf{x})\,,
\end{equation}
Then, let us focus on eq. (\ref{eq:NRQCD}) in the sector in which heavy quarks with momentum of order $\sqrt{MT}$ interact with gluons with energy of order $T$.  This would be equivalent to performing the matching between NRQCD and $\textrm{NRQCD}_{sh}$ at tree level
\begin{equation}
\mathcal{L}_{\xi}=\sum_{\mathbf{p}\neq 0}\xi^\dagger_\mathbf{p}\left\{iD_0-\frac{(\mathbf{p}-i\boldsymbol{\nabla}+g\mathbf{A})^2}{2M}+c_4\frac{p^4}{8M^3}+c_Fg\frac{\boldsymbol{\sigma}\cdot\mathbf{B}}{2M}\right\}\xi_\mathbf{p}+\cdots
\end{equation}
It is more convenient to rearrange the terms in the following way
\begin{equation}
\mathcal{L}_\xi=\sum_{\mathbf{p}\neq 0}\xi^\dagger_\mathbf{p}\left\{iD_0-\frac{p^2}{2M}+i\frac{\mathbf{p}\cdot(\boldsymbol{\nabla}+ig\mathbf{A})}{M}+\frac{(\boldsymbol{\nabla}+ig\mathbf{A})^2}{2M}+c_4\frac{p^4}{8M^3}+c_Fg\frac{\boldsymbol{\sigma}\cdot\mathbf{B}}{2M}\right\}\xi_\mathbf{p}+\cdots
\label{eq:lxi}
\end{equation}
The advantage of this equation compared to the previous one is that now each term has a well-defined power counting. Since the only scale that has not been integrated out is $T$, it follows that $\xi_\mathbf{p}$ is of size $T^{3/2}$. Therefore, the first two terms in eq. (\ref{eq:lxi}) are of order $1$. The third term is of order $\sqrt{\frac{T}{M}}$ and the rest of terms are of order $\frac{T}{M}$. We have obtained eq. (\ref{eq:lxi}) by tree-level manipulations of the NRQCD Lagrangian. However, it is easy to convince our selves that it corresponds to the heavy quark sector of $\textrm{NRQCD}_{sh}$. The only differences that a proper matching would bring up in this case are possible sub-leading corrections to $M$, and the Wilson coefficients $c_4$ and $c_F$. Note that Galilean symmetry and reparametrization invariance \cite{Luke:1992cs} constraint the form of eq. (\ref{eq:lxi}).

Let us now mention some features of $\textrm{NRQCD}_{sh}$ we believe are worth emphasizing. The first one is that, up to a trivial shift in the energy, the field $\xi_\mathbf{p}$ behaves at leading-order as a static quark. This implies that we can use $\textrm{NRQCD}_{sh}$ to compute properties of a heavy quark with a momentum of order $\sqrt{MT}$ by performing perturbations around the static case. The second remarkable feature that we want to comment is that the study of the Wigner distribution is very much simplified in $\textrm{NRQCD}_{sh}$. The Wigner distribution of a heavy quark in NRQCD takes the following form
\begin{equation}
f_W(\mathbf{p},\mathbf{R})=\int\,d^3r\mathrm{e}^{-i\mathbf{p}\mathbf{r}}\mathit{Tr}\left(\psi^\dagger\left(t,\mathbf{R}-\frac{\mathbf{r}}{2}\right)\phi\left(t:\mathbf{R}-\frac{\mathbf{r}}{2},\mathbf{R}+\frac{\mathbf{r}}{2}\right)\psi\left(t,\mathbf{R}+\frac{\mathbf{r}}{2}\right)\rho\right)\,,
\label{eq:deffw}
\end{equation}
where $\phi$ is a Wilson line introduced in the definition such that the Wigner distribution is gauge invariant \cite{Vasak:1987um} and $\rho$ is the density matrix. By virtue of this Wilson line, the $\mathbf{p}$ momentum in the Wigner distribution corresponds to the kinetic momentum and not to the canonical momentum conjugate \cite{Vasak:1987um}. We are interested in the case in which $p\sim\sqrt{MT}$ and the dependence of $f_W$ with $R$ is only sizeable over distances much larger than $1/T$. In this case, $f_W$ is matched in $\textrm{NRQCD}_{sh}$ into
\begin{equation}
\begin{split}
 f_W(\mathbf{r},\mathbf{R})=Z_0(\mathbf{r})\mathit{Tr}\left(\xi^\dagger_\mathbf{p}(t,\mathbf{R})\xi_\mathbf{p}(t,\mathbf{R})\rho\right)\\
 +iZ_1(\mathbf{r})\mathit{Tr}\left(\mathbf{r}\cdot\mathbf{D}\xi^\dagger_\mathbf{p}(t,\mathbf{R})\xi_\mathbf{p}(t,\mathbf{R})-\mathbf{r}\cdot\xi^\dagger_\mathbf{p}(t,\mathbf{R})\mathbf{D}\xi_\mathbf{p}(t,\mathbf{R})\rho\right)\\
 +Z_{2}(\mathbf{r})r\mathit{Tr}\left(\xi^\dagger_\mathbf{p}(t,\mathbf{R})\mathbf{r}\cdot\mathbf{E}(t,\mathbf{R})\xi_\mathbf{p}(t,\mathbf{R})\rho\right)+\cdots
 \end{split}
 \end{equation} 
 where $f_W(\mathbf{r},\mathbf{R})$ is the Fourier transform of eq. (\ref{eq:deffw}). This equation is valid up to terms smaller than $\frac{T}{M}$. Note that, since both the lhs and the rhs can be understood as a pseudo-probability distribution normalized to $1$, $Z_0(\mathbf{r})=1+\mathcal{O}(r^2T^2)$. In conclusion, up to sub-leading corrections, we can identify $\mathit{Tr}\left(\xi^\dagger_\mathbf{p}(t,\mathbf{R})\xi_\mathbf{p}(t,\mathbf{R})\rho\right)$ with the Wigner distribution. The use of $\textrm{NRQCD}_{sh}$ allows seeing the Wigner distribution as a probability distribution (at least at leading order) encoded in a local operator.
 
Let us now discuss some properties of the propagators in $\textrm{NRQCD}_{sh}$ in thermal field theory that will be useful in the following. Now on, and in order to simplify the notation, we will drop the sub-index $\mathbf{p}$ from the field $\xi$. First, let us discuss the dilute limit, which we define as the limit in which
\begin{equation}
-\mathit{Tr}(\xi(t,\mathbf{r}_1)\xi^\dagger(t,\mathbf{r}_2)\rho)\gg\mathit{Tr}(\xi^\dagger(t,\mathbf{r}_2)\xi(t,\mathbf{r}_1)\rho)\,.
\end{equation}
In this limit, 
\begin{equation}
S_{11}(t,\mathbf{r})=\mathit{Tr}(\mathcal{P}\xi_1(t,\mathbf{r})\xi_1^\dagger(0,\mathbf{0})\rho)\sim\theta(t)\mathit{Tr}(\xi(t,\mathbf{r})\xi^\dagger(0,\mathbf{0})\rho)\,,
\end{equation}
\begin{equation}
S_{22}(t,\mathbf{r})=\mathit{Tr}(\mathcal{P}\xi_2(t,\mathbf{r})\xi_2^\dagger(0,\mathbf{0})\rho)\sim\theta(-t)\mathit{Tr}(\xi(t,\mathbf{r})\xi^\dagger(0,\mathbf{0})\rho)\,,
\end{equation}
\begin{equation}
S_{12}(t,\mathbf{r})=S^<(t,\mathbf{r})=\mathit{Tr}(\mathcal{P}\xi_1(t,\mathbf{r})\xi_2^\dagger(0,\mathbf{0})\rho)=-\mathit{Tr}(\xi^\dagger(0,\mathbf{0})\xi(t,\mathbf{r}))\rho)\sim 0\,,
\end{equation}
and
\begin{equation}
S_{21}(t,\mathbf{r})=S^>(t,\mathbf{r})=\mathit{Tr}(\mathcal{P}\xi_2(t,\mathbf{r})\xi_1^\dagger(0,\mathbf{0})\rho)\sim\mathit{Tr}(\xi(t,\mathbf{r})\xi^\dagger(0,\mathbf{0})\rho)\,.
\end{equation}
We note that these results are true as long as we are exactly in the dilute limit. It is also useful to write the expression in the Keldysh representation \cite{Keldysh:1964ud}. 
\begin{equation}
S_R(t,\mathbf{r})=\theta(t)\mathit{Tr}(\{\xi(t,\mathbf{r})\xi^\dagger(0,\mathbf{0})\}\rho)\sim S_{11}(t,\mathbf{r})\,,
\end{equation}
\begin{equation}
S_A(t,\mathbf{r})=-\theta(-t)\mathit{Tr}(\{\xi(t,\mathbf{r})\xi^\dagger(0,\mathbf{0})\}\rho)\sim -S_{22}(t,\mathbf{r})\,,
\end{equation}
and 
\begin{equation}
S_S(t,\mathbf{r})=\mathit{Tr}([\xi(t,\mathbf{r})\xi^\dagger(0,\mathbf{0})]\rho)\sim S_{21}(t,\mathbf{r})\,.
\end{equation}
The following useful relation is exactly fulfilled in the dilute limit
\begin{equation}
S_S(t,\mathbf{r})\sim S_R(t,\mathbf{r})-S_A(t,\mathbf{r})\,.
\end{equation}
This implies that in the dilute limit the retarded propagator contains all the relevant information, since the advanced propagator is the complex conjugate of the retarded.

Let us now consider that the density of heavy quarks is small but non-zero. In this case, $S^<$ is small but not zero. Since this propagator is directly related with the distribution of heavy quarks, finite density corrections in other propagators have a sub-leading impact on the mentioned distribution.

Finally, let us write the heavy quark propagator in $\textrm{NRQCD}_{sh}$ at tree level. 
The tree level propagators are more compactly written in the Keldysh representation
\begin{equation}
S^0_R(t,\mathbf{r})=\int\frac{\,d^4k}{(2\pi)^4}\mathrm{e}^{-ik_0t+i\mathbf{k}\mathbf{r}}\frac{i}{k_0-\frac{p^2}{2M}+i\epsilon}=\theta(t)\mathrm{e}^{-\frac{ip^2t}{2M}}\delta^{(3)}(\mathbf{r})\,,
\end{equation}
\begin{equation}
S^0_A(t,\mathbf{r})=\int\frac{\,d^4k}{(2\pi)^4}\mathrm{e}^{-ik_0t+i\mathbf{k}\mathbf{r}}\frac{i}{k_0-\frac{p^2}{2M}-i\epsilon}=-\theta(-t)\mathrm{e}^{-\frac{ip^2t}{2M}}\delta^{(3)}(\mathbf{r})\,,
\end{equation}
and
\begin{equation}
\begin{split}
S^0_S(t,\mathbf{r},\mathbf{R})=\mathit{Tr}([\xi^0\left(t,\mathbf{R}+\frac{\mathbf{r}}{2}\right){\xi^{0}}^{\dagger}\left(0,\mathbf{R}-\frac{\mathbf{r}}{2}\right)]\rho)\\
=\int\frac{\,d^4k}{(2\pi)^4}\mathrm{e}^{-ik_0t+i\mathbf{k}\mathbf{r}}2\pi\delta\left(k_0-\frac{p^2}{2M}\right)(1-2f^0(\mathbf{p}+\mathbf{k},\mathbf{R}))\,,
\label{eq:fdef}
\end{split}
\end{equation}
where the $0$ super-index denotes tree level quantities and we have assumed that the tree level distribution function of heavy quarks $f^0$ is a very smooth function over distances of size $\mathbf{r}$. Note also that to be consistent with the $\textrm{NRQCD}_{sh}$ power counting, the function $f^0$ (or the resummed equivalent that we will introduce later) must be expanded
\begin{equation}
 f^0(\mathbf{p}+\mathbf{k},\mathbf{R})\sim f^0(\mathbf{p},\mathbf{R})+\mathbf{k}\boldsymbol{\nabla}_\mathbf{p}f^0(\mathbf{p},\mathbf{R})+\frac{k^ik^j}{2}\Delta^{ij}_\mathbf{p}f^0(\mathbf{p},\mathbf{R})+\cdots
\label{eq:disexp}
 \end{equation} 
This implies that
\begin{equation}
\begin{split}
S^0_S(t,\mathbf{r},\mathbf{R})\sim\mathrm{e}^{-\frac{ip^2t}{2M}}\left[(1-2f^0(\mathbf{p},\mathbf{R}))\delta^{(3)}(\mathbf{r})+2\boldsymbol{\nabla}\delta^{(3)}(\mathbf{r})\boldsymbol{\nabla}_\mathbf{p}f^0(\mathbf{p},\mathbf{R})\right.\\
\left.-\Delta^{ij}\delta^{(3)}(\mathbf{r})\Delta^{ij}_\mathbf{p}f^0(\mathbf{p},\mathbf{R})+\cdots\right]\,.
\end{split}
\label{eq:fexpansion}
\end{equation}
Finally, let us mention that the Feynmann rules for $\textrm{NRQCD}_{sh}$ can be found in Appendix \ref{sec:frnrqcdl}.
\section{Langevin Effective Theory}
\label{sec:let}
In this section we introduce Langevin Effective Theory (LET). It is obtained from $\textrm{NRQCD}_{sh}$ after integrating out degrees of freedom with an energy of the order of the scale $T$. In this case we are dealing with heavy quarks with momentum $\mathbf{p}+\mathbf{k}$  where now the residual momentum $k$ is much smaller than the temperature. Note that we can always redefine $p$ such that this relation is fulfilled. In order to construct LET, we follow the observations of \cite{Baidya:2017eho}. The influence functional of LET  can be written as $\mathrm{e}^{iS_{LET}}$, with
\begin{equation}
S_{LET}=S_1-S_2+\mathcal{I}+S_{HTL}\,,
\label{eq:let1}
\end{equation}
where $S_1$ ($S_2$) is the piece of the action that involves only heavy quark fields of type $1$ ($2$). $\mathcal{I}$ is a new type of contribution in which heavy quark fields of both type $1$ and $2$ appear. Finally, $S_{HTL}$ is the well-known Hard Thermal Loop action \cite{Braaten:1991gm}. If the scale $T$ did not induce any dissipative effects, then $\mathcal{I}$ would be zero and $S_2$ would be equal to $S_1$ just changing the fields of type $2$ by fields of type $1$.

 The construction of LET is substantially simplified by applying the dilute limit. We know that, in this limit, we can ignore the doubling of degrees of freedom when computing Green functions in which only heavy fields of type $1$ (or $2$) appear \cite{Escobedo:2008sy,Brambilla:2008cx}. Moreover, in the dilute limit and as long as the heavy particle is non-relativistic in the frame in which the medium is at rest, the symmetries of $\textrm{NRQCD}_{sh}$ are not broken by the presence of the medium. Therefore, the form of $S_1$ is equal to the action of $\textrm{NRQCD}_{sh}$. However, the Wilson coefficients are different. In fact, some of these Wilson coefficients can be complex. Then, 
 \begin{equation}
 \begin{split}
 S_1=\int\,d^4x\xi^\dagger_1\left[(1-\delta Z)\left(iD_0-\frac{p^2}{2M}\right)-\delta E+i\frac{\Gamma}{2}+i\frac{\mathbf{p}\cdot(\boldsymbol{\nabla}+ig\mathbf{A})}{M}+\frac{(\boldsymbol{\nabla}+ig\mathbf{A})^2}{2M}\right.\\
 \left.+\tilde{c}_4\frac{p^4}{8M^3}+\tilde{c}_Fg\frac{\boldsymbol{\sigma}\cdot\mathbf{B}}{2M}\right]\xi_1+\cdots
 \end{split}
 \label{eq:S1}
 \end{equation}
 where we allow both $\delta Z$, $\delta E$ and $\Gamma$ to be polynomials involving $M$, $T$ and $p^2$. Note that $\delta Z$ can have both a real and an imaginary part. The real part of $\delta Z$ can be reabsorbed by a unitary transformation. Whether the imaginary part of $\delta Z$ can be reabsorbed by a transformation of the fields is beyond the scope of this work. We note that, since our main focus is the study of $\frac{df}{dt}$, we can ignore a non-zero value of $\delta Z$. The reason is that the wave-function renormalization is not a secular effect, meaning that its effects on the evolution on the distribution of heavy quarks does not become larger as we study longer times. This is in contrast to what happens to corrections to the mass and the decay width. For example, even if the decay width is small, it becomes a leading order effect at large enough times. 

$S_2$ can be obtained from $S_1$ by changing the fields of type $1$ to fields of type $2$ and by making the complex conjugate of the Wilson coefficients. In our case, this means making the changes $\delta Z\to\delta Z^*$ and $i\frac{\Gamma}{2}\to -i\frac{\Gamma}{2}$. We note also, that since $\delta E$ and $\Gamma$ are obtained by performing a matching computation to $\textrm{NRQCD}_{sh}$, they are polynomials with the following structure
\begin{equation}
\delta E=\alpha_0 T+\alpha_1\frac{T^2}{M}+\alpha_2\frac{p^2T}{M^2}+\mathcal{O}\left(\frac{T^3}{M^2}\right)\,,
\label{eq:expdE}
\end{equation}
\begin{equation}
\Gamma=\beta_0 T+\beta_1\frac{T^2}{M}+\beta_2\frac{p^2T}{M^2}+\mathcal{O}\left(\frac{T^3}{M^2}\right)\,.
\label{eq:expG}
\end{equation}

Regarding $\mathcal{I}$, it can be fixed by imposing the following conditions:
\begin{itemize}
\item $S_{LET}$ is equal to zero if fields of type $1$ are equal to fields of type $2$ \cite{Baidya:2017eho}.
\item In the dilute limit, the propagator $S_{12}$ is zero.
\end{itemize}
Using this, we get
\begin{equation}
\mathcal{I}=\int\,d^4x\xi^\dagger_2\left[2i\mathrm{Im} Z\left(iD_0-\frac{p^2}{2M}\right)-i\Gamma-i\mathrm{Im}\tilde{c}_4\frac{p^4}{4M^3}-i\mathrm{Im}\tilde{c}_F\frac{\boldsymbol{\sigma}\cdot\mathbf{B}}{M}\right]\xi_1+\cdots
\label{eq:I}
\end{equation}
Let us now discuss the issue of gauge invariance. The equations we wrote are only invariant regarding transformations in which the fields are modified in the same in both branches of the Schwinger-Keldysh contour. More specifically, there is an explicit invariance under the following type of infinitesimal gauge transformations:
\begin{equation}
    \begin{split}
        \xi_1\to \xi_1+ig\Lambda \xi_1\,, \\
        \xi_2\to \xi_2+ig\Lambda \xi_2\,, \\
        A^\mu_1\to A^\mu_1+\partial^\mu\Lambda+ig[\Lambda,A^\mu_1]\,, \\
        A^\mu_2\to A^\mu_2+\partial^\mu\Lambda+ig[\Lambda,A^\mu_2]\,.
    \end{split}
\end{equation}
Apparently, something has been missed in going from $\textrm{NRQCD}_{sh}$ to LET. $\textrm{NRQCD}_{sh}$ is invariant under transformations in which each branch of the Schwinger-Keldysh contour is independently transformed with a different $\Lambda_i$. How to recover this more general invariance in EFTs where terms mixing the two branches appear was discussed in \cite{Haehl:2016pec,Crossley:2015evo,Jensen:2017kzi}, however the solution the found is not suitable for our case. We have proposed an alternative solution in appendix \ref{sec:gauge}, where we discuss this issue and its solution in more detail. However, at the end of the day, this issue has little practical importance for the computation at hand and will only complicate the notation. Careful readers might have notived that we also did not mention whether the gauge fields entering $D_0$ and $\mathbf{B}$ in eq. (\ref{eq:I}) are of type $1$ or $2$. More details about this are also given in appendix \ref{sec:gauge}.

The previous way of presenting the action of $S_{LET}$ is useful to perform the matching to the full theory. However, we might rearrange the contributions in a more physically meaningful way.
\begin{equation}
S_{LET}=\tilde{S}_1-\tilde{S}_2+\mathcal{D}\,,
\end{equation}
where
 \begin{equation}
 \begin{split}
 \tilde{S}_i=\int\,d^4x\xi^\dagger_i\left[(1-\mathrm{Re}\delta Z)\left(iD_0-\frac{p^2}{2M}\right)-\delta E+i\frac{\mathbf{p}\cdot(\boldsymbol{\nabla}+ig\mathbf{A})}{M}+\frac{(\boldsymbol{\nabla}+ig\mathbf{A})^2}{2M}\right.\\
 \left.+\mathrm{Re}\tilde{c}_4\frac{p^4}{8M^3}+\mathrm{Re}\tilde{c}_Fg\frac{\boldsymbol{\sigma}\cdot\mathbf{B}}{2M}\right]\xi_i+\cdots
 \end{split}
 \end{equation}
 is the unitary part of the evolution and $\mathcal{D}$ is the dissipative part
\begin{equation}
\begin{split}
\mathcal{D}=i\int\,d^4x\left(
\begin{array}{cc}
\xi^\dagger_1 & \xi^\dagger_2
\end{array}
\right)\left[-\mathrm{Im}\delta Z\left(iD_0-\frac{p^2}{2M}\right)+\frac{\Gamma}{2}+\mathrm{Im}\tilde{c}_4\frac{p^4}{8M^3}+\mathrm{Im}\tilde{c}_F\frac{\boldsymbol{\sigma}\cdot\mathbf{B}}{2M}\right]\\
\otimes\left(
\begin{array}{cc}
1 & 0 \\
-2 & 1
\end{array}
\right)\left(
\begin{array}{c}
\xi_1 \\
\xi_2
\end{array}\right)\,.
\end{split}
\end{equation}

It might be illustrative to write $S_{LET}$ in the Keldysh basis \cite{Keldysh:1964ud,Chou:1984es}. Let us introduce
\begin{equation}
\xi_{Cl}=\frac{\xi_1+\xi_2}{2}\,,
\end{equation}
which we call the \textit{classical} field and
\begin{equation}
\xi_Q=\xi_1-\xi_2\,,
\label{eq:defq}
\end{equation}
the \textit{quantum} field. There are two remarkable properties that can be seen just introducing this basis. First, the condition that $S_{LET}$ is zero when fields of type $1$ are equal of type $2$ can be rephrased as imposing that $S_{LET}=0$ when $\xi_{Q}=0$. It also follows that a QFT describing a closed system (without dissipation) only has terms containing an odd number of \textit{quantum} fields. However, in an EFT obtained after integrating out medium degrees of freedom (such as the one we are studying in this paper), we might have terms containing an even number of \textit{quantum} fields. In summary, terms with an even number of \textit{quantum} fields are forbidden in $\tilde{S}_i$ while they are allowed in $\mathcal{D}$. To see this more explicitly, let us introduce the following operators:
\begin{equation}
    \begin{split}
    \Delta_S=(1-\mathrm{Re}\delta Z)\left(iD_0-\frac{p^2}{2M}\right)-\delta E+i\frac{\mathbf{p}\cdot(\boldsymbol{\nabla}+ig\mathbf{A})}{M}+\frac{(\boldsymbol{\nabla}+ig\mathbf{A})^2}{2M}\\
    +\mathrm{Re}\tilde{c}_4\frac{p^4}{8M^3}+\mathrm{Re}\tilde{c}_Fg\frac{\boldsymbol{\sigma}\cdot\mathbf{B}}{2M}\,,
    \end{split}
\end{equation}
and
\begin{equation}
        \Delta_D=-\mathrm{Im}\delta Z\left(iD_0-\frac{p^2}{2M}\right)+\frac{\Gamma}{2}+\mathrm{Im}\tilde{c}_4\frac{p^4}{8M^3}+\mathrm{Im}\tilde{c}_F\frac{\boldsymbol{\sigma}\cdot\mathbf{B}}{2M}\,.
\end{equation}
Then, we can write $S_{LET}$ in the following way
\begin{equation}
    S_{LET}=\int\,d^4x\left(\xi^\dagger_{Cl}\Delta_S\xi_Q+\xi^\dagger_Q\Delta_S\xi_{Cl}+\xi^\dagger_Q\Delta_D\xi_{Cl}-\xi^\dagger_{Cl}\Delta_D\xi_Q+\xi^\dagger_Q\Delta_D\xi_Q\right)\,.
    \label{eq:letqclb}
\end{equation}
Writing $S_{LET}$ in this way provides some extra insight. In a situation in which $\xi_Q$ is suppressed, the leading contribution comes from terms linear in $\xi_Q$. Integrating over $\xi_Q$ considering only these linear terms gives a Dirac delta that forces $\xi_{Cl}$ to follow the classical equations of motion. Terms quadratic in $\xi_Q$ can be seen as originating from a classical random source, therefore we can understand them as fluctuations \cite{Greiner:1998vd}.

Now, let us discuss the case in which we still consider that heavy quarks are dilute but we take into account the first corrections proportional to their density. We call this NLO dilute corrections. 
Regarding the symmetries of the EFT, we consider that $f(\mathbf{p})$ does not have any preferred direction other than $\mathbf{p}$ itself. 
Taking this into account, $\mathcal{D}$ has to be modified in order to include an extra term
\begin{equation}
\begin{split}
\mathcal{D}=i\int\,d^4x\left(
\begin{array}{cc}
\xi^\dagger_1 & \xi^\dagger_2
\end{array}
\right)\left[\left(-\mathrm{Im}\delta Z\left(iD_0-\frac{p^2}{2M}\right)+\frac{\Gamma}{2}+\mathrm{Im}\tilde{c}_4\frac{p^4}{8M^3}+\mathrm{Im}\tilde{c}_F\frac{\boldsymbol{\sigma}\cdot\mathbf{B}}{2M}\right)\right.\\
\left.\otimes\left(
\begin{array}{cc}
1 & 0 \\
-2 & 1
\end{array}
\right)
+\Delta\Gamma\left(
\begin{array}{cc}
0 & -1 \\
1 & 0
\end{array}\right)\right]\left(
\begin{array}{c}
\xi_1 \\
\xi_2
\end{array}\right)\,.
\end{split}
\label{eq:Dnlo}
\end{equation}
We note that the term that we have added is the leading one that we can add that does not fulfill the condition that the $S_{12}$ propagator has to be zero (dilute limit) but that fulfills the rest of conditions that we have discussed in this section. Again, we can rewrite the action in the \textit{classical}-\textit{quantum} basis,
\begin{equation}
    \begin{split}
        S_{LET}=\int\,d^4x\left(\xi^\dagger_{Cl}\Delta_S\xi_Q+\xi^\dagger_Q\Delta_S\xi_{Cl}+\xi^\dagger_Q\Delta_D\xi_{Cl}-\xi^\dagger_{Cl}\Delta_D\xi_Q+\xi^\dagger_Q\Delta_D\xi_Q\right.\\
        \left.+\xi^\dagger_{Cl}\Delta\Gamma\xi_Q-\xi^\dagger_Q\Delta\Gamma\xi_{Cl}\right)\,.
    \end{split}
    \label{eq:ClQ}
\end{equation}

We note that all Wilson coefficients can be affected by the NLO dilute corrections in a sub-leading way. The specific property of $\Delta\Gamma$ is that it vanishes in the exact dilute limit.

There are additional symmetries in the EFT that impose relations between the different terms in $\Gamma$ and $\Delta\Gamma$. The origin of these relations is the Schwinger-Keldysh symmetry \cite{Haehl:2016pec,Crossley:2015evo,Jensen:2017kzi}, also known as the fluctuation-dissipation theorem. However, let us postpone this discussion until next section, since it is very much related with the evolution of $f(\mathbf{p})$.

\section{Evolution of $f(\mathbf{p})$ in LET}
\label{sec:finlet}
In this section, we show how to compute the evolution of $f(\mathbf{p})$ in the EFT we introduced before. At the same time, this will also allow us to introduce further constraints on the Wilson coefficients by imposing that the fluctuation-dissipation theorem is fulfilled\footnote{$f$ is a function of $\mathbf{p}$ and $\mathbf{R}$. However, we assume that $f$ is a very smooth function in $R$ when we look at distances of the order of $1/T$. Therefore, we can consider that $f$ does not depend on for the purposes of the matching and the study in this section. We also assume that the distribution is isotropic.}. On more physical terms, this means the following. We are studying the case of a dilute distribution of heavy quarks evolving in a large bath in thermal equilibrium at a temperature such that $T\ll M$. Therefore, at very large times $f(\mathbf{p})$ must be equal to the thermal distribution, $\mathrm{e}^{-\frac{p^2}{2MT}}$.

The information on the distribution of heavy quarks can be found more directly in the $S^<(p)$ propagator, which, as we have seen before, goes to zero in the dilute limit. In order to be more precise, we define the distribution function $f$ such that 
\begin{equation}
    S^<(k)=f(\mathbf{p}+\mathbf{k})(S_R(k)-S_A(k))\,,
\end{equation}
where, as usual, we are refering to the propagators of the field $\xi_p$. To study this propagator, it is convenient to use the Kadanoff-Baym equations \cite{Baym:1961zz}. A recent application of these equations in the context of heavy-ion collisions can be found in \cite{Sheng:2021kfc}. These equations are deduced by performing a Dyson-Schwinger type of resummation of the self-energies. In our case, we will just perform a resummation of the tree-level self-energies as obtained directly from the LET influence functional.  
\begin{equation}
    \begin{split}
        \Pi_R=\delta E-i\frac{\Gamma}{2}\,, \\
        \Pi_A=\delta E+i\frac{\Gamma}{2}\,, \\
        \Pi^<=-i\Delta\Gamma\,.                
    \end{split}
\end{equation}
The first and second lines of the previous equation follow directly from eq. (\ref{eq:letqclb}). The third line follows from eq. (\ref{eq:Dnlo}). We note that, when studying a system evolving in a plasma, some kind of resummation is always needed to deal with secular effects. In other words, small perturbations that grow with time need to be resummed because they become leading order effects at large enough times.

The propagator $S^<$ is a function of two times, 
\begin{equation}
    S^<(t_1,\mathbf{r}_1;t_2,\mathbf{r}_2)=\textit{Tr}(\xi_1(t_1,\mathbf{r}_1)\xi^\dagger(t_2,\mathbf{r}_2)\rho)\,.
\end{equation}
In thermal equilibrium and due to translational invariance it is only a function of $\tau=t_1-t_2$. More generally, it is also a function of $t=\frac{t_1+t_2}{2}$. In order to study the evolution of $f(\mathbf{p})$, the more direct way is to look at the evolution of $S^<$ as a function of $t$ for $\tau=0$. The Kadanoff-Baym equations give us the evolution on each time separately
\begin{equation}
    \begin{split}
        \partial_{t_1}S^<=-i\left(\frac{p^2}{2M}+\Pi_R\right)S^<+i\Pi^<S_A\,, \\
        \partial_{t_2}S^<=i\left(\frac{p^2}{2M}+\Pi_A\right)S^<-i\Pi^<S_R\,. \\
    \end{split}
\end{equation}
From this, it follows that the evolution with $t$ is given by
\begin{equation}
    \partial_t S^<=-i(\Pi_R-\Pi_A)S^<-i\Pi^<(S_R-S_A)\,.
    \label{eq:dS}
\end{equation}
Note that the spectral function $\rho$ is given by $\rho=S_R-S_A$, which at tree level is $\rho_0=2\pi\delta\left(k_0-\frac{p^2}{2M}\right)$. We are studying the case of a heavy particle interacting with a medium in thermal equilibrium. At very large times we should arrive to a steady state in which the distribution of heavy particles is also in thermal equilibrium. This means that at late times
\begin{equation}
    S^<=-\frac{\Pi^<(S_R-S_A)}{\Pi_R-\Pi_A}\,.
\end{equation}
At the same time, at thermal equilibrium the fluctuation-dissipation theorem must be fulfilled
\begin{equation}
    S^<=f_{eq}(\mathbf{p}+\mathbf{k})(S_R-S_A)\,,
    \label{eq:rSf} 
\end{equation}
where $f_{eq}(p)=N\mathrm{e}^{-\frac{p^2}{2M}}$ is the thermal equilibrium distribution function and $N$ a normalization factor. We obtain the relation
\begin{equation}
    \Pi^<(\mathbf{p},\mathbf{k},f_{eq})=f_{eq}(\mathbf{p}+\mathbf{k})(\Pi_R(\mathbf{p},\mathbf{k})-\Pi_A(\mathbf{p},\mathbf{k}))\,,
\end{equation}
that must be fulfilled. Note that we have explicitly written the dependency with the momentum and the distribution function. To simplify, let us focus on the case $\mathbf{k}=\mathbf{0}$ and write the expression in terms of the Wilson coefficients of LET
\begin{equation}
    \Delta\Gamma\left(\mathbf{p},\mathrm{e}^{-\frac{p^2}{2M}}\right)=\mathrm{e}^{-\frac{p^2}{2M}}\Gamma(\mathbf{p})\,.
    \label{eq:relDGamma}
\end{equation}
$\Delta\Gamma$ is proportional to $f$ or its derivatives
\begin{equation}
    \Delta\Gamma(\mathbf{p},f_{eq})=\left(\gamma^0+\frac{\gamma^1}{M}+\frac{\gamma^2p^2}{M^2}\right)f_{eq}(\mathbf{p})+\frac{\gamma^3}{M}\mathbf{p}\nabla_\mathbf{p}(f_{eq}(\mathbf{p}))+\gamma^4\Delta_\mathbf{p}f_{eq}(\mathbf{p})+\mathcal{O}\left(\frac{T}{M}\right)^{3/2}\,.
\end{equation}
At the same time, a similar expansion is valid for $\Gamma$
\begin{equation}
    \Gamma=\beta_0+\frac{\beta_1}{M}+\frac{\beta_2p^2}{M^2}\,.
    \label{eq:Gbeta}
\end{equation}
Then, this implies that the following relations must be fulfilled:
\begin{equation}
    \begin{split}
        \gamma_0=\beta_0\,,\\
        \gamma_1-\frac{3\gamma_4}{T}=\beta_1\,,\\
        \gamma_2-\frac{\gamma_3}{T}+\frac{\gamma_4}{T^2}=\beta_2\,.
    \end{split}
    \label{eq:rel}
\end{equation}
In order to compute $\frac{\partial f(\mathbf{p})}{\partial t}$, we can use eqs. (\ref{eq:dS}), (\ref{eq:rSf}) and (\ref{eq:rel}) in the case $\mathbf{k}=\mathbf{0}$
\begin{equation}
    \frac{\partial f(\mathbf{p})}{\partial t}=-\Gamma f(\mathbf{p})+\Delta\Gamma f(\mathbf{p})=\left(\frac{3}{M}-\frac{p^2}{M^2T}\right)\left(\frac{\gamma_4}{T}-\gamma_3\right)f(\mathbf{p})+\frac{\gamma_3}{M}\mathbf{\nabla}(\mathbf{p}f(\mathbf{p}))+\gamma_4\Delta_\mathbf{p}f(\mathbf{p})\,.
\end{equation}
Another condition that must be fulfilled is unitarity. This implies that 
\begin{equation}
    \frac{\partial}{\partial t}\int\,d^3pf(\mathbf{p})=0\,.
\end{equation}
This imposes an additional condition, this is $\gamma_4=T\gamma_3$. Note that what we obtain is the evolution of $f(\mathbf{p})$ that corresponds to a Fokker-Planck equation, in which the drag coefficient is given by $\frac{\gamma_3}{M}$.

In summary, imposing all the constraints coming from unitarity and the fluctuation-dissipation theorem, we get the following expression for $\Delta\Gamma$
\begin{equation}
    \Delta\Gamma(\mathbf{p},f)=\left(\beta_0+\frac{\beta_1}{M}+\frac{\beta_2p^2}{M^2}\right)f(\mathbf{p})+\frac{\kappa}{2MT}\nabla_\mathbf{p}(\mathbf{p}f(\mathbf{p}))+\frac{\kappa}{2}\Delta_\mathbf{p}f(\mathbf{p})+\mathcal{O}\left(\frac{T}{M}\right)^{3/2}\,,
    \label{eq:DGamma}
\end{equation}
where $\kappa$ is the heavy quark diffusion coefficient \cite{Casalderrey-Solana:2006fio}. We see that the matching between $\mathrm{NRQCD}_{sh}$ and $LET$ gets substantially simplified. To determine $\Delta \Gamma$ at the order we are interested, we only need to know $\Gamma$ in the dilute limit and the value of the heavy quark diffusion coefficient. Moreover, if we are only interested in the evolution of $f(\mathbf{p})$ we only need to know $\kappa$, as it was expected, and we can see by writing the evolution of $f(\mathbf{p})$ considering all the constraints
\begin{equation}
    \frac{\partial f(\mathbf{p})}{\partial t}=\frac{\kappa}{2MT}\mathbf{\nabla}(\mathbf{p}f(\mathbf{p}))+\frac{\kappa}{2}\Delta_\mathbf{p}f(\mathbf{p})\,.
    \label{eq:Lanfinal}
\end{equation}
\section{Matching between $\mathrm{NRQCD}_{sh}$ and LET}
\label{sec:matching}
In this section, we are going to perform the matching between $\mathrm{NRQCD}_{sh}$ and LET in the one gluon exchange approximation. The reader might wonder why we do not compute the matching at a given order in perturbation theory instead. The reason is related to the special properties of gauge theories at finite temperature. A one-loop matching will lead to $\kappa=0$, meaning that $f(\mathbf{p})$ does not change. On the other hand, a two-loop matching would involve a complex computation of $\Gamma$ that has little phenomenological impact due to the symmetries that we discussed in the previous section. The one gluon exchange approximation has the virtue of giving a finite contribution to all Wilson coefficients already at leading order, and we can regard it as an intermediate step between a one-loop and a two-loop matching. 

The strategy to perform the matching is going to be the following. First, we are going to compute $\Pi_R$ in $\mathrm{NRQCD}_{sh}$, from this we can obtain $\delta M$ and $\Gamma$. Second, we are going to compute $\Pi^<$, but only the piece proportional to $\mathbf{\nabla}_\mathbf{p}f(\mathbf{p})$, since we have seen in section \ref{sec:finlet} that this is enough to fix all the relevant Wilson coefficients. Regarding the computation of $\Pi_R$, we are going to use the Keldysh basis following the graphical notation of \cite{Ghiglieri:2020dpq}\footnote{Propagators with a single arrow pointing to the right (left) are retarded (advanced). Propagators with two outgoing arrows and a \textit{capacitor} symbol are symmetric. The rest of the symbols are defined in appendix \ref{sec:frnrqcdl}}. For the computation of $\Pi^<$, we find it more convenient to use the $1-2$ basis. We note that in this section we are going to use the Feynmann rules discussed in Appendix \ref{sec:frnrqcdl}.

\begin{figure}
    \begin{minipage}{0.45\textwidth}
        \centering
        \begin{tikzpicture}
            \draw[->] (0,0) -- (1,0);
            \draw (1,0) -- (2.5,0);
            \draw[->] (2.9,0) -- (2.5,0);
            \draw (2.9,-0.25) -- (2.9,0.25);
            \draw (3.1,-0.25) -- (3.1,0.25);
            \draw[dashed,->] (2,0) arc (180:90:1);
            \draw[dashed] (4,0) arc (0:90:1);
            \draw[->] (3.1,0) -- (3.5,0);
            \draw[->] (3.5,0) -- (5,0);
            \draw (5,0) -- (6,0);
        \end{tikzpicture}
    \end{minipage}
    \hfill
    \begin{minipage}{0.45\textwidth}
        \centering
        \begin{tikzpicture}
            \coordinate (A) at ({3-cos(45)},{sin(45)});
            \draw[->] (0,0) -- (1,0);
            \draw[->] (1,0) -- (3,0);
            \draw[->] (3,0) -- (5,0);
            \draw (5,0) -- (6,0);
            \draw[dashed] (2,0) arc (180:90+45:1);
            \draw[dashed] (4,0) arc (0:45:1);
            \draw[dashed,<-] (A) arc (90+45:95:1);
            \draw[dashed,<-] ({3+cos(45)},{sin(45)}) arc (45:85:1);
            \draw (2.9,0.75) -- (2.9,1.25);
            \draw (3.1,0.75) -- (3.1,1.25);
        \end{tikzpicture}
    \end{minipage}
\caption{Diagrams contribution to $\Pi_R$ at order $T$ in $\mathrm{NRQCD}_{sh}$. The computation is done in the Keldysh basis.}
\label{fig:LOPiR}
\end{figure}
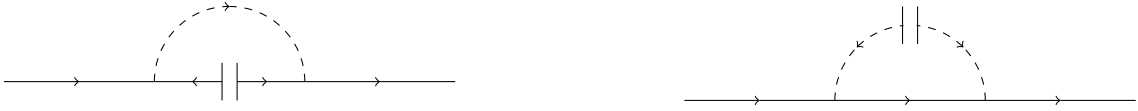

Let us begin by computing the leading contribution to $\Pi_R$ in $\mathrm{NRQCD}_{sh}$, in other words, the terms that in the EFT power counting are of order $T$. In fig. \ref{fig:LOPiR} we show the relevant diagrams. If we write $\delta E$ as
\begin{equation}
    \delta E=\alpha_0+\frac{\alpha_1}{M}+\frac{\alpha_2p^2}{M^2}\,,
\end{equation}
and we use the notation of eq. (\ref{eq:Gbeta}), then the contribution from the diagrams in fig. \ref{fig:LOPiR} is 
\begin{equation}
    \begin{split}
        \alpha_0=-\frac{ig^2C_F}{4}\int\frac{\,d^3q}{(2\pi)^3}(\Delta_{00,R}+\Delta_{00,A})(0,\mathbf{q})\,, \\
        \beta_0=2Tg^2C_F\int\frac{\,d^3q}{(2\pi)^3}\rho_{00}'(0,\mathbf{q})\,,
    \end{split}
\end{equation}
where $\Delta_{00}$ is the temporal gluon propagator and $\rho'$ is the first derivative over the temporal component of the spectral function of the temporal gluon propagator \cite{Bellac:2011kqa}. Our computation is performed in the Coulomb gauge.

\begin{figure}
    \begin{minipage}{0.45\textwidth}
        \begin{tikzpicture}
            \coordinate (A) at ({8./3.},0);
            \draw[->] (0,0) -- (1,0);
            \draw (1,0) -- ({(7./3.)},0);
            \draw[->] (3,0) -- ({7./3.},0);
            \filldraw[white] (A) circle (3pt);
            \draw (A) circle (3pt);
            \draw[->] ({10./3.-0.1},0) -- (3,0);
            \draw ({10./3.-0.1},-0.25) -- ({10./3.-0.1},0.25);
            \draw ({10./3.+0.1},-0.25) -- ({10./3.+0.1},0.25);
            \draw[dashed,->] (2,0) arc (180:90:1);
            \draw[dashed] (4,0) arc (0:90:1);
            \draw[->] ({10./3.+0.1},0) -- ({11./3.},0);
            \draw[->] ({11./3.},0) -- (5,0);
            \draw (5,0) -- (6,0);
        \end{tikzpicture}
    \end{minipage}
    \hfill
    \begin{minipage}{0.45\textwidth}
        \begin{tikzpicture}
            \coordinate (A) at ({10./3.},0);
            \draw[->] (0,0) -- (1,0);
            \draw (1,0) -- ({(7./3.)},0);
            \draw[->] ({8./3.-0.1},0) -- ({7./3.},0);
            \draw[->] ({8./3.+0.1},0) -- (3,0);
            \draw ({8./3.-0.1},-0.25) -- ({8./3.-0.1},0.25);
            \draw ({8./3.+0.1},-0.25) -- ({8./3.+0.1},0.25);
            \draw[->] (3,0) -- ({11./3.},0);
            \filldraw[white] (A) circle (3pt);
            \draw (A) circle (3pt);
            \draw[dashed,->] (2,0) arc (180:90:1);
            \draw[dashed] (4,0) arc (0:90:1);
            \draw[->] ({11./3.},0) -- (5,0);
            \draw (5,0) -- (6,0);
        \end{tikzpicture}
    \end{minipage}
    \centering
    \begin{tikzpicture}
        \coordinate (A) at ({3-cos(45)},{sin(45)});
        \draw[->] (0,0) -- (1,0);
        \draw[->] (1,0) -- (2.5,0);
        \draw[->] (2.5,0) -- (3.5,0);
        \draw[->] (3.5,0) -- (5,0);
        \filldraw[white] (3,0) circle (3pt);
        \draw (3,0) circle (3pt);
        \draw (5,0) -- (6,0);
        \draw[dashed] (2,0) arc (180:90+45:1);
        \draw[dashed] (4,0) arc (0:45:1);
        \draw[dashed,<-] (A) arc (90+45:95:1);
        \draw[dashed,<-] ({3+cos(45)},{sin(45)}) arc (45:85:1);
        \draw (2.9,0.75) -- (2.9,1.25);
        \draw (3.1,0.75) -- (3.1,1.25);
    \end{tikzpicture}
    \caption{Diagrams contribution to $\Pi_R$ at order $T\sqrt{\frac{T}{M}}$ in $\mathrm{NRQCD}_{sh}$.}
\label{fig:NLOPiR}
\end{figure}
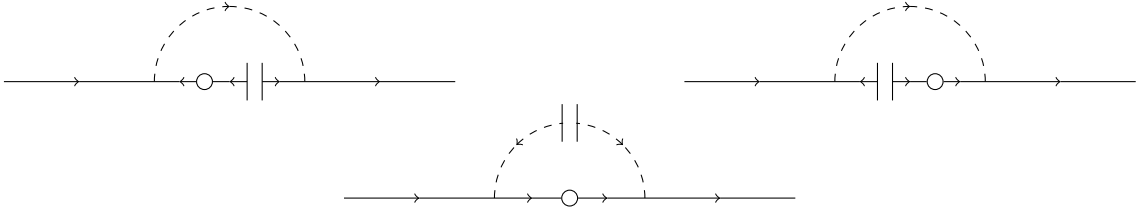

Next, let us discuss possible corrections to $\delta E$ and $\Gamma$ of order $T\sqrt{\frac{T}{M}}$. By symmetry arguments, there should not be any. However, let us check explicitly that it is the case. The relevant diagrams are shown in fig. \ref{fig:NLOPiR}. Each circled correction to the propagator introduces a factor $\frac{\mathbf{p}\mathbf{q}}{M}$, where $\mathbf{q}$ is the spatial momentum of the temporal gluon. We can take $\mathbf{p}$ out of the integrand, and since the medium does not have any preferred direction in space, the result of the integration can only be zero.

\begin{figure}
    \begin{minipage}{0.45\textwidth}
        \begin{tikzpicture}
            \coordinate (A) at ({8./3.},0);
            \draw[->] (0,0) -- (1,0);
            \draw (1,0) -- ({(7./3.)},0);
            \draw[->] (3,0) -- ({7./3.},0);
            \filldraw[white] (A) circle (3pt);
            \draw ({8./3.-0.1},-0.1) rectangle ++(0.2,0.2);
            \draw[->] ({10./3.-0.1},0) -- (3,0);
            \draw ({10./3.-0.1},-0.25) -- ({10./3.-0.1},0.25);
            \draw ({10./3.+0.1},-0.25) -- ({10./3.+0.1},0.25);
            \draw[dashed,->] (2,0) arc (180:90:1);
            \draw[dashed] (4,0) arc (0:90:1);
            \draw[->] ({10./3.+0.1},0) -- ({11./3.},0);
            \draw[->] ({11./3.},0) -- (5,0);
            \draw (5,0) -- (6,0);
        \end{tikzpicture}
    \end{minipage}
    \hfill
    \begin{minipage}{0.45\textwidth}
        \begin{tikzpicture}
            \coordinate (A) at ({10./3.},0);
            \draw[->] (0,0) -- (1,0);
            \draw (1,0) -- ({(7./3.)},0);
            \draw[->] ({8./3.-0.1},0) -- ({7./3.},0);
            \draw[->] ({8./3.+0.1},0) -- (3,0);
            \draw ({8./3.-0.1},-0.25) -- ({8./3.-0.1},0.25);
            \draw ({8./3.+0.1},-0.25) -- ({8./3.+0.1},0.25);
            \draw[->] (3,0) -- ({11./3.},0);
            \filldraw[white] (A) circle (3pt);
            \draw ({10./3.-0.1},-0.1) rectangle ++(0.2,0.2);
            \draw[dashed,->] (2,0) arc (180:90:1);
            \draw[dashed] (4,0) arc (0:90:1);
            \draw[->] ({11./3.},0) -- (5,0);
            \draw (5,0) -- (6,0);
        \end{tikzpicture}
    \end{minipage}
    \begin{minipage}{0.45\textwidth}
        \begin{tikzpicture}
            \coordinate (A) at ({3-cos(45)},{sin(45)});
            \draw[->] (0,0) -- (1,0);
            \draw[->] (1,0) -- (2.5,0);
            \draw[->] (2.5,0) -- (3.5,0);
            \draw[->] (3.5,0) -- (5,0);
            \filldraw[white] (3,0) circle (3pt);
            \draw (2.9,-0.1) rectangle ++(0.2,0.2);
            \draw (5,0) -- (6,0);
            \draw[dashed] (2,0) arc (180:90+45:1);
            \draw[dashed] (4,0) arc (0:45:1);
            \draw[dashed,<-] (A) arc (90+45:95:1);
            \draw[dashed,<-] ({3+cos(45)},{sin(45)}) arc (45:85:1);
            \draw (2.9,0.75) -- (2.9,1.25);
            \draw (3.1,0.75) -- (3.1,1.25);
        \end{tikzpicture}
    \end{minipage}
    \hfill
    \begin{minipage}{0.45\textwidth}
        \begin{tikzpicture}
            \draw[->] (0,0) -- (1,0);
            \draw[->] (1,0) -- (3,0);
            \draw (3,0) -- (4,0);
            \draw[gluon] (2,0) arc (-90:270:0.5);
            \filldraw[white] (2,0) circle (3pt);
            \filldraw[white] (1.9,0.75) rectangle ++(0.2,0.5);
            \draw (1.9,0.75) -- (1.9,1.25);
            \draw (2.1,0.75) -- (2.1,1.25);
            \draw(1.9,-0.1) rectangle ++(0.2,0.2);
            \draw[->] (2.37,0.5) -- ++(0,-0.005);
            \draw[->] (1.62,0.4) -- ++(0,-0.005);
        \end{tikzpicture}
    \end{minipage}
    \caption{Diagrams contribution to $\alpha_1$ and $\beta_1$ in $\mathrm{NRQCD}_{sh}$.}
\label{fig:NNLOPiRa}
\end{figure}
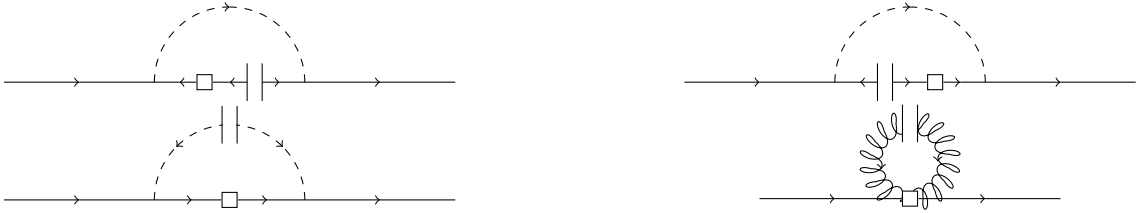

Regarding the diagrams contributing at order $\frac{T^2}{M}$, we can divide them in two classes. Those that contribute to $\alpha_1$ and $\beta_1$, and those contributing to $\alpha_2$ and $\beta_2$. The first class is shown in fig. \ref{fig:NNLOPiRa}. They give the following contribution:
\begin{equation}
    \begin{split}
        \alpha_1=\frac{g^2C_F}{4}\int\frac{\,d^4q}{(2\pi)^4}\Delta_{ii,S}(q)+\frac{g^2C_F}{8}\int\frac{\,d^4q}{(2\pi)^4}q^2\left(\frac{1}{(q_0-i\epsilon)^2}+\frac{1}{(q_0+i\epsilon)^2}\right)\Delta_{00,S}(q) \\
        -\frac{ig^2C_F}{8}\int\frac{\,d^3q}{(2\pi)^3}q^2\frac{d(\Delta^R_{00}+\Delta^A_{00})}{dq_0}(0,\mathbf{q})\,,\\
        \beta_1=-\frac{g^2C_f}{4}\int\frac{\,d^3q}{(2\pi)^3}q^2\rho'_{00}(0,\mathbf{q})\,.
    \end{split}
\end{equation}

\begin{figure}
    \begin{minipage}{0.45\textwidth}
        \begin{tikzpicture}
            \draw[->] (0,0) -- (1,0);
            \draw[->] (1,0) -- (3,0);
            \draw[->] (3,0) -- (5,0);
            \draw (5,0) -- (6,0);
            \draw[gluon] (2,0) arc (180:0:1);
            \filldraw[white] (1.9,-0.1) rectangle ++(0.2,0.2);
            \draw (1.9,-0.1) rectangle ++(0.2,0.2);
            \filldraw[white] (3.9,-0.1) rectangle ++(0.2,0.2);
            \draw (3.9,-0.1) rectangle ++(0.2,0.2);
            \filldraw[white] (2.9,0.75) rectangle ++(0.2,0.5);
            \draw (2.9,0.75) -- (2.9,1.25);
            \draw (3.1,0.75) -- (3.1,1.25);
            \draw[->] ({2.9-cos(45.)},{0.1+sin(45.)}) -- ++(-0.005,-0.005);
            \draw[->] ({3.15+cos(45.)},{sin(45.)}) -- ++(0.00,-0.005);
        \end{tikzpicture}
    \end{minipage}
    \hfill
    \begin{minipage}{0.45\textwidth}
        \begin{tikzpicture}
            \draw[->] (0,0) -- (1,0);
            \draw (1,0) -- (2.5,0);
            \draw[<->] (2.5,0) -- (3.5,0);
            \draw[->] (3.5,0) -- (5,0);
            \draw (5,0) -- (6,0);
            \condensat{3}{0}
            \draw[gluon] (2,0) arc (180:0:1);
            \quadrat{2}{0}
            \quadrat{4}{0}
            \fletxa{3}{1.15}{0}
        \end{tikzpicture}        
    \end{minipage}
    \begin{minipage}{0.45\textwidth}
        \begin{tikzpicture}
            \draw[->] (0,0) -- (1,0);
            \draw[->] (1,0) -- ({7./3.},0);
            \draw[->] ({7./3.},0) -- (3,0);
            \draw[->] (3,0) -- ({11./3.},0);
            \draw[->] ({11./3.},0) -- (5,0);
            \draw (5,0) -- (6,0);
            \rodona{{8./3.}}{0}
            \rodona{{10./3.}}{0}
            \draw[dashed] (2,0) arc (180:0:1);
            \condensat{3}{1}
            \fletxa{{3.+cos(45.)}}{{sin(45.)}}{-45.}
            \fletxa{{3.-cos(45.)}}{{sin(45.)}}{225.}
        \end{tikzpicture}
    \end{minipage}
    \hfill
    \begin{minipage}{0.45\textwidth}
        \begin{tikzpicture}
            \draw[->] (0,0) -- (1,0);
            \draw (1,0) -- (2.25,0);
            \draw[<->] (2.25,0) -- (2.75,0);
            \condensat{2.5}{0}
            \draw[->] (2.75,0) -- (3.25,0);
            \draw[->] (3.25,0) -- (3.75,0);
            \draw[->] (3.75,0) -- (5,0);
            \draw (5,0) -- (6,0);
            \rodona{3}{0}
            \rodona{3.5}{0}
            \draw[dashed] (2,0) arc (180:0:1);
            \fletxa{3}{1}{0}
        \end{tikzpicture}        
    \end{minipage}
    \begin{minipage}{0.45\textwidth}
        \begin{tikzpicture}
            \draw[->] (0,0) -- (1,0);
            \draw (1,0) -- (2.25,0);
            \draw[->] (2.75,0) -- (2.25,0);
            \draw[<->] (2.75,0) -- (3.25,0);
            \draw[->] (3.25,0) -- (3.75,0);
            \draw[->] (3.75,0) -- (5,0);
            \draw (5,0) -- (6,0);
            \rodona{2.5}{0}
            \condensat{3}{0}
            \rodona{3.5}{0}
            \draw[dashed] (2,0) arc (180:0:1);
            \fletxa{3}{1}{0}
        \end{tikzpicture}
    \end{minipage}
    \hfill
    \begin{minipage}{0.45\textwidth}
        \begin{tikzpicture}
            \draw[->] (0,0) -- (1,0);
            \draw[->] (1,0) -- (2.25,0);
            \draw[->] (2.25,0) -- (2.75,0);
            \draw (2.75,0) -- (3.25,0);
            \draw[<->] (3.25,0) -- (3.75,0);
            \draw[->] (3.75,0) -- (5,0);
            \draw (5,0) -- (6,0);
            \rodona{3}{0}
            \rodona{2.5}{0}
            \condensat{3.5}{0}
            \draw[dashed] (2,0) arc (180:0:1);
            \fletxa{3}{1}{0} 
        \end{tikzpicture}        
    \end{minipage}
    \caption{Diagrams contribution to $\alpha_2$ and $\beta_2$ in $\mathrm{NRQCD}_{sh}$.}
\label{fig:NNLOPiRb}
\end{figure}
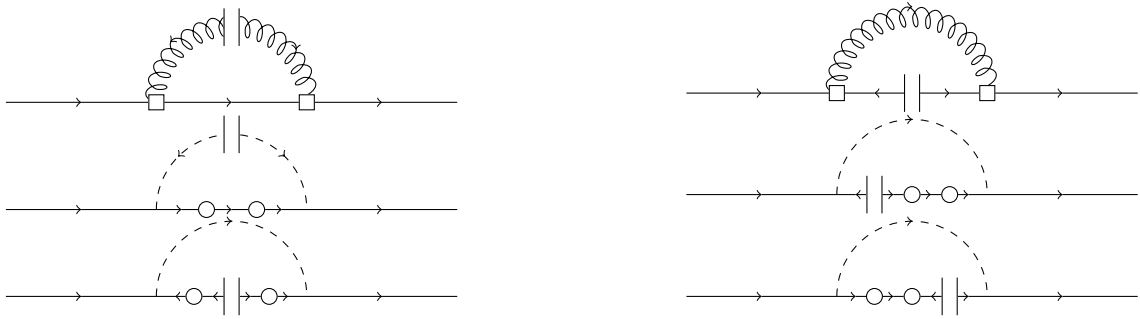
Diagrams contributing to $\alpha_2$ and $\beta_2$ are shown in fig. \ref{fig:NNLOPiRb}. They result in the following contribution:
\begin{equation}
    \begin{split}
        \alpha_2=-i\frac{g^2C_F}{12}\int\frac{\,d^3q}{(2\pi)^3}(\Delta_{ii,R}+\Delta_{ii,A})\,, \\
        \beta_2=\frac{g^2C_FT}{6}\int\frac{\,d^3q}{(2\pi)^3}\rho_{ii}'(0,\mathbf{q})-\frac{g^2}{6}\int\frac{\,d^4q}{(2\pi)^4}\left(\frac{i}{q_0+i\epsilon}\right)^3\Delta_{00,S}(q)\mathbf{q}^2\,.
    \end{split}
\end{equation}

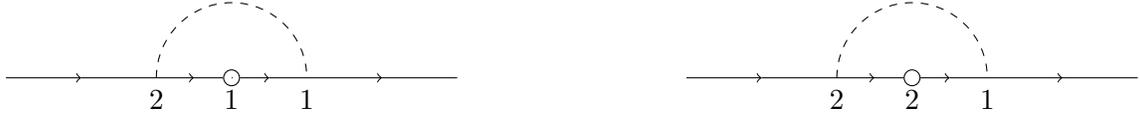
\begin{figure}
    \begin{minipage}{.45\textwidth}
        \begin{tikzpicture}
            \draw[->] (0,0) -- (1,0);
            \draw[->] (1,0) -- (2.5,0);
            \draw[->] (2.5,0) -- (3.5,0);
            \draw[->] (3.5,0) -- (5,0);
            \draw (5,0) -- (6,0);
            \draw[dashed] (2,0) node[anchor=north, yshift=-1] {2} arc (180:0:1);
            \draw (4,0) node[anchor=north, yshift=-1] {1} -- ++(0.00001,0);
            \rodona{3}{0}
            \draw (3,0) node[anchor=north, yshift=-1] {1} -- ++(0.00001,0);
        \end{tikzpicture}
    \end{minipage}
    \hfill
    \begin{minipage}{.45\textwidth}
        \begin{tikzpicture}
            \draw[->] (0,0) -- (1,0);
            \draw[->] (1,0) -- (2.5,0);
            \draw[->] (2.5,0) -- (3.5,0);
            \draw[->] (3.5,0) -- (5,0);
            \draw (5,0) -- (6,0);
            \draw[dashed] (2,0) node[anchor=north, yshift=-1] {2} arc (180:0:1);
            \draw (4,0) node[anchor=north, yshift=-1] {1} -- ++(0.00001,0);
            \rodona{3}{0}
            \draw (3,0) node[anchor=north, yshift=-1] {2} -- ++(0.00001,0);
        \end{tikzpicture}
    \end{minipage}
    \caption{Diagrams involved in the matching of $\kappa$.}
    \label{fig:drag}
\end{figure}

Let us now compute $\Pi^<$ in $\mathrm{NRQCD}_{sh}$ in order to perform the matching. We are going to focus on the piece proportional to $\mathbf{p}\mathbf{\nabla}_\mathbf{p}f(\mathbf{p})$ since, as we discussed previously, it is the only extra term needed to perform the matching. To obtain this, we need to go to second order in the expansion shown in eq. (\ref{eq:fexpansion}). Moreover, we have to take from the $\mathrm{NRQCD}_{sh}$ Lagrangian the interaction that goes like $\frac{\xi^\dagger\mathbf{p}\mathbf{\nabla}\xi}{M}$. Therefore, we need to focus on the diagram shown in fig. \ref{fig:drag}. Note that for the computation of $\Pi^<$ we are using the $1-2$ basis instead of the Schwinger-Keldysh one. From this matching, we obtain
\begin{equation}
    \kappa=\frac{g^2C_F T}{3}\int\frac{\,d^3q}{(2\pi)^3}q^2\rho'_{00}(0,\mathbf{q})\,.
    \label{eq:kappapert}
\end{equation}
We can check that this result is compatible with the perturbative computation of $\kappa$ in the Coulomb gauge \cite{Casalderrey-Solana:2006fio,Caron-Huot:2009ncn,Moore:2004tg}. Note that in the $q_0\to 0$ limit we can approximate $\mathbf{E}$ by $-\mathbf{\nabla}A_0$, as long as we are not using the unitary gauge. It is also important to take into account that we are making a small abuse of language by identifying the transport coefficient $\kappa$ with the Wilson coefficient of LET. Strictly speaking, the Wilson coefficient only includes the contribution to $\kappa$ from the scale we are integrating out, in this case $T$. 

\begin{figure}
    \begin{center}
    \begin{tikzpicture}
        \draw[->] (0,0) -- (1,0);
        \draw (0,0) -- (2,0) node[anchor=north] {2};
        \draw[->] (2,0) -- (3,0);
        \draw (3,0) -- (4,0) node[anchor=north] {1};
        \draw[->] (4,0) -- (5,0);
        \draw (5,0) -- (6,0);
        \draw[dashed] (2,0) arc (180:0:1);
    \end{tikzpicture}
\end{center}
\caption{An alternative diagram to consider in order to match $\kappa$.}
\label{fig:dragalt}
\end{figure}
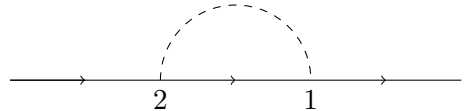

With this we have finished the matching to the order of interest. However, for the sake of cross-checking some of the relations we discussed in the previous section, we are going to discuss an alternative way to match the value of $\kappa$. This can be done looking at the diagram in fig. \ref{fig:dragalt}. However, now we have to take the third term in eq. (\ref{eq:fexpansion}) to get the contribution proportional to $\Delta f(\mathbf{p})$. Doing this, we can check that we get exactly the same value of $\kappa$. There is even a third way of computing $\kappa$ which is to look at the terms in $\Delta \Gamma$ that go like $\frac{f(\mathbf{p})}{M}$ and subtract $\beta^1$. However, we are not going to discuss further this way of doing the matching of $\kappa$.

Now we are ready to discuss how to derive a Fokker-Planck equation for heavy quarks in QCD including the effects of the scales $T$ and $gT$. The first step has already been discussed. We match $\mathrm{NRQCD}_{sh}$ to LET to include the effect of the scale $T$ in the Wilson coefficients. The second step is to compute $\Pi_R$ and $\Pi^<$ in LET including the effects of the scale $gT$. Note however that the same symmetries and cancellations apply now, and therefore, we only need to compute the contribution of the scale $gT$ to $\kappa$. This can be obtained by computing the diagram in fig. \ref{fig:drag}, but now the loop has to be computed using the HTL gluon propagator. Proceeding in this way, we obtain the perturbative estimate of $\kappa$. It corresponds to eq. (\ref{eq:kappapert}) but now using the two loop perturbative result for $q\sim T$ and the HTL approximation for $q\sim m_D$ \cite{Caron-Huot:2009ncn,Moore:2004tg}. 

\section{Heavy quarks with momentum $p\sim T$}
\label{sec:psimT}
In this section, we discuss the case of a heavy quark with a tri-momentum of order $T$. In this case, we will proceed as follows:
\begin{itemize}
    \item As starting point, we will use NRQCD Lagrangian.
    \item We will integrate out the scale $T$. After this we will arrive again to LET. However, now the Wilson coefficients and the power counting will be slightly different, since now $p$ is of order $T$ instead of $\sqrt{MT}$ as before. 
    \item In this case, $\Delta\Gamma$ will have a more involved dependency with $p$. As a result, instead of a Fokker-Planck equation, we get an evolution similar to a Boltzmann equation.
\end{itemize}
Let us start looking at LET. The Lagrangian is again formed by combining eq. (\ref{eq:S1}) and eq. (\ref{eq:I}) in the same way as before. 
The main difference is the value of the Wilson coefficients. 

Regarding $\delta E$ and $\Gamma$, it is easy to check that the results are exactly the same as in the case $p\sim\sqrt{MT}$. The reason is that the diagrams that appear doing the respective matchings in $\mathrm{NRQCD}_{sh}$ (for $p\sim\sqrt{MT}$) and in NRQCD (for $p\sim T$) are the same at any order in the coupling constant, the only difference being the relative size of each term. However, the situation is different for $\Delta\Gamma$, the reason is that now this Wilson coefficient is, in general, a non-trivial function of $f$ and $p$ and we can not assume a polynomial form like in eq. (\ref{eq:DGamma}). However, we can assume that it is a polynomial in inverse powers of $1/M$ from the structure of NRQCD. Then, we can write
\begin{equation}
    \Delta\Gamma(\mathbf{p},f)=\int\frac{\,d^3q}{(2\pi)^3}\left(C_1(q)+\frac{C_2(q)q^2}{M}+\frac{C_3(q)(2\mathbf{q}\mathbf{p}-q^2)}{M}\right)f(\mathbf{p}-\mathbf{q})\,.
\end{equation}
There are several constraints that can be imposed to $C_i$. Eq. (\ref{eq:relDGamma}) must be fulfilled, but now we need to take into account that we need to expand $\mathrm{e}^{-\frac{p^2}{2MT}}$ to the order desired since $\frac{p^2}{2MT}\ll 1$. This implies that
\begin{equation}
    \begin{split}
        \int\frac{\,d^3q}{(2\pi)^3}C_1(q)=\beta_0\,, \\
        \int\frac{\,d^3q}{(2\pi)^3}q^2\left(C_2(q)-C_3(q)-\frac{C_1(q)}{2T}\right)=\beta_1\,, 
    \end{split}
    \label{eq:Boltzcond}
\end{equation}
Additional constraints come from unitarity. We can repeat the arguments of section \ref{sec:finlet} to arrive to 
\begin{equation}
    \partial_t f(\mathbf{p})=\Delta\Gamma(\mathbf{p},f)-\Gamma f(\mathbf{p})\,.
\end{equation}
Imposing that $\partial_t\int\frac{\,d^3p}{(2\pi)^3}f(\mathbf{p}=0)$ we deduce that
\begin{equation}
    \int\frac{\,d^3p}{(2\pi)^3}\Delta\Gamma(\mathbf{p},f)=\Gamma\,.
\end{equation}
Starting from the previous equation and using the change of variables $\mathbf{p}\to \mathbf{p}+\mathbf{q}$ we obtain the additional constraint
\begin{equation}
    \int\frac{\,d^3q}{(2\pi)^3}q^2\left(C_2(q)+C_3(q)\right)=\beta_1\,.
    \label{eq:Boltzcond2}
\end{equation}
Using the previous equation and some trivial manipulations, we arrive at our final form of the evolution equation
\begin{equation}
    \begin{split}
    \partial_t f(\mathbf{p})=\int\frac{\,d^3q}{(2\pi)^3}\left(C_1(q)+\frac{C_2(q)q^2}{M}\right)(f(\mathbf{p}-\mathbf{q})-f(\mathbf{p}))\\
    +\frac{1}{M}\int\frac{\,d^3q}{(2\pi)^3}C_3(q)(2\mathbf{q}\mathbf{p}-q^2)(f(\mathbf{p}-\mathbf{q})+f(\mathbf{p}))\,.
    \end{split}
    \label{eq:Boltzmann}
\end{equation}

As an illustration, let us discuss the perturbative matching. This can be done by looking at the diagrams in figs. \ref{fig:dragalt} and \ref{fig:drag}, but now taking into account that the expansion in eq. (\ref{eq:fexpansion}) can not be done since $p\sim q$. We obtain that
\begin{equation}
    \begin{split}
        C_1(q)=g^2C_FT\rho'_{00}(0,\mathbf{q})\,, \\
        C_3(q)=-\frac{g^2C_F\rho'_{00}(0,\mathbf{q})}{4}\,, \\
        C_2(q)=0\,.
    \end{split}
\end{equation}
Actually, the last equation seems to be true at any perturbative order just because of the structure of NRQCD. We can see explicitly that the conditions in eqs. (\ref{eq:Boltzcond}) and (\ref{eq:Boltzcond2}) are fulfilled.

Finally, let us discuss the physical picture that the EFT treatment of the cases $p\sim T$ and $p\sim \sqrt{MT}$ provides when they are put together. If we start with a heavy quark at rest with the medium, it will start to gain momentum due to collisions. At this stage, the evolution of the system is given by a Boltzmann equation like eq. (\ref{eq:Boltzmann}). This evolution is such that there is almost an equal probability to gain momentum than to lose it. Therefore, we will naturally arrive to a situation in which $p\gg T$ after some time. In this case, the evolution can already be described by a Langevin equation. As the momentum keeps increasing, the drag force starts to become important. Finally, we arrive to an equilibrium distribution in which momenta much larger than $\sqrt{MT}$ are very rare. Let us note that the arguments needed to arrive to these qualitative statements are valid even if the medium is strongly coupled.  

\section{Discussion and conclusions}
\label{sec:concl}
 In this manuscript, we have discussed an EFT approach to the derivation of the evolution of a heavy quark in a medium. Our main motivation was to pave the way for future developments in the study of quarkonium suppression. We would like to obtain a Lindblad equation for quarkonium without assuming a weakly-coupled plasma and using the hierarchy of energy scales that appears in the problem. This has already been done in the case that the medium sees quarkonium as a small color dipole \cite{Brambilla:2016wgg,Brambilla:2017zei}. However, we would like to generalize to the case $Tr\sim 1$ as this would allow modeling more realistically excited states of bottomonium and charmonium at LHC energies. This is challenging because the EFT in which heavy quarkonium is best described is pNRQCD, and it is obtained after integrating out the scale $1/r$. However, if $Tr\sim 1$ the matching between NRQCD and pNRQCD is modified by the medium. The appearance of an imaginary part of the potential indicates that the medium induces dissipative effects that can not be encoded in an effective action, instead we need an influence functional in which terms mixing the two branches of the Schwinger-Keldysh contour appear. Therefore, we need to modify pNRQCD to include this kind of terms. The motivation of this manuscript is to study a simpler case in order to pave the way for this modification of pNRQCD in the future.

 The simpler case we studied is a single heavy quark interacting with a medium. We did it following the EFT philosophy, in which each energy scale must be treated separately using a series of EFTs. We distinguished two cases, the case $p\sim \sqrt{MT}$ and the case $p\sim T$. In both cases we start with NRQCD, however, in the first case we integrate out the scale $\sqrt{MT}$ as an intermediate step going from NRQCD to $\textrm{NRQCD}_{sh}$. Finally, we integrate the scale $T$ in both cases to go to a new EFT that we named LET. The structure of this EFT is given by eq. (\ref{eq:ClQ}), which is one of the main results of this manuscript. This formula combines the constraints of the Keldysh symmetry together with the special features of heavy particles in the dilute limit. Further constraints are provided by the fluctuation-dissipation theorem, which together with the EFT power counting leads to eq. (\ref{eq:Lanfinal}) for $p\sim\sqrt{MT}$ and to eq. (\ref{eq:Boltzmann}) for $p\sim T$. This is the second main result from this manuscript, that eqs. (\ref{eq:Lanfinal}) and (\ref{eq:Boltzmann}) can be obtained simply by symmetry and power counting arguments within the EFT approach. The physical picture is also quite transparent, a heavy quark with momentum of order $T$ can be described by a Boltzmann equation and will slowly increase its momentum until we reach a moment in  which the momentum is large enough so that a Langevin equation becomes valid. 

Let us now discuss in more details the relevance of the study made in this manuscript in the context of quarkonium suppression. We have seen that the state of a heavy quark can not be described by a single position $x$, instead we need the position in the upper(lower) Schwinger-Keldysh branch $x_1$, ($x_2$). Note that $x_1-x_2$ is related to $p$ by the Wigner transform. The increase of $p$ we discussed below is seen in coordinate space as a narrowing of the $x_1-x_2$ coordinate or, put differently, the density matrix becoming almost diagonal in the coordinate basis. This behavior has also been observed in quarkonium \cite{Blaizot:2017ypk,Delorme:2021eno}. One main difference between the case of quarkonium and that of a single heavy quark is that the number of coordinates needed to describe the system doubles. In quarkonium, we can talk about the center-of-mass coordinate, $R$, and the relative coordinate, $r$. However, in non-equilibrium situations we need to consider the doubling of degrees of freedom, and then we have to take into account four coordinates, $R_1$, $R_2$, $r_1$ and $r_2$. A complete EFT treatment of quarkonium would identify all the possible scale hierarchies that can be constructed with these coordinates and that appear in the problem of quarkonium suppression. The case studied in this manuscript corresponds to a situation in which the quark and the antiquark are very far apart, and we have studied the cases $x_1-x_2\sim\frac{x_1+x_2}{2}$ and $x_1-x_2\ll\frac{x_1+x_2}{2}$. In \cite{Brambilla:2016wgg,Brambilla:2017zei}, the case studied corresponds to $r_1-r_2\sim\frac{r_1+r_2}{2}\ll\frac{1}{T}$ and $T\gg E$. Perturbative studies have shown that the Lindblad equation that appears in the case $T\gg E$ naturally leads to a decrease of $r_1-r_2$. This produces a change in the power counting that, among other things, makes the drag force not a perturbation (NLO corrections in $E/T$ that include the drag force were added in \cite{Brambilla:2022ynh,Brambilla:2023hkw}). Regarding the relation between the scales $1/r$, $T$ and $E$; pNRQCD has been studied for all possible relations \cite{Escobedo:2008sy,Brambilla:2010vq,Brambilla:2008cx} but only taking into account the analogous to $S_i$ part in the notation of eq. (\ref{eq:let1}). This is enough to discuss spectroscopy, but not to study the evolution of quarkonium inside a plasma. What is missing, and we hope this work helps to develop, is the construction of the analogous to $\mathcal{I}$ in pNRQCD. 

Finally, let us discuss possible extensions of the present work in the context of heavy quark propagation itself. As in any EFT, an obvious improvement would be to compute higher order corrections in the $\sqrt{\frac{T}{M}}$ expansion to improve our knowledge about the evolution of $f(\mathbf{p})$. Another possible direction is to relax the assumptions about the symmetries of the problem. For example, until now we have assumed that both the medium and the distribution of heavy quarks are homogeneous in space and isotropic. It would be interesting to relax these conditions, as they are not completely fulfilled in heavy-ion collisions. Finally, the case $p\gg \sqrt{MT}$ is also interesting. In this case, we would be studying the case of a heavy quark that loses energy until thermalizing with the medium.

\acknowledgments{MAE wants to thanks the careful reading of a first version of this manuscript and the useful comments of Juan Torres-Rincon, Joan Soto, Nora Brambilla and Antonio Vairo. The work of MAE has been supported by the Maria de Maetzu excellence program under project CEX2019-000918-M, by the Spanish Research State Agency under project PID2019-105614GB-C21 and by the grant 2021-SGR-249 of Generalitat de Catalunya.}

\appendix
\section{Feynmann rules for $\textrm{NRQCD}_{sh}$}
\label{sec:frnrqcdl}
The Lagrangian of the heavy quark sector of $\textrm{NRQCD}_{sh}$ is
\begin{equation}
\mathcal{L}_\xi=\sum_{\mathbf{p}\neq 0}\xi^\dagger_\mathbf{p}\{iD_0-\frac{p^2}{2M}+i\frac{\mathbf{p}\cdot(\boldsymbol{\nabla}+ig\mathbf{A})}{M}+\frac{(\boldsymbol{\nabla}+ig\mathbf{A})^2}{2M}+c_4\frac{p^4}{8M^3}+c_Fg\frac{\boldsymbol{\sigma}\cdot\mathbf{B}}{2M}\}\xi_\mathbf{p}+\cdots
\end{equation}
now on we will not write explicit the sub-script $\mathbf{p}$ in this section as this quantity is conserved in all diagrams. They Feynmann rules for this theory are
\begin{equation}
\begin{tikzpicture}
\draw[->] (0,0) -- (1,0);
\draw (1,0) -- (2,0);
\end{tikzpicture}=\frac{i}{k_0-\frac{p^2}{2M}+i\epsilon}\,,
\end{equation}
\begin{equation}
\begin{tikzpicture}[baseline={([yshift=-.5ex]current bounding box.center)}]
\draw[->] (0,0) node [anchor=north] {$\mathbf{k}$} -- (0.5,0);
\draw[->] (0.5,0) -- (1.5,0);
\draw (1.5,0) -- (2,0) node [anchor=north] {$\mathbf{k}+\mathbf{q}$};
\draw[dashed] (1,1) node [anchor=west] {$\mathbf{q}$} -- (1,0);
\end{tikzpicture}=ig\,,
\label{eq:prop}
\end{equation}
\begin{equation}
\begin{tikzpicture}[baseline={([yshift=-.5ex]current bounding box.center)}]
\draw[->] (0,0) node [anchor=north] {$\mathbf{k}$} -- (0.5,0);
\draw[->] (0.5,0) -- (1.5,0);
\draw (1.5,0) -- (2,0) node [anchor=north] {$\mathbf{k}+\mathbf{q}$};
\draw[gluon] (1,1) node [anchor=west] {$\mathbf{q},i$} -- (1,0);
\end{tikzpicture}=-\frac{igp^i}{M}\,,
\end{equation}
\begin{equation}
\begin{tikzpicture}
\draw[->] (0,0) -- (0.5,0);
\draw[->] (0.5,0) -- (1.5,0);
\draw (1.5,0) -- (2,0);
\filldraw[white] (1,0) circle (3pt);
\draw (1,0) circle (3pt);
\end{tikzpicture}=-\frac{i\mathbf{p}\cdot\mathbf{k}}{M}\,,
\end{equation}
\begin{equation}
\begin{tikzpicture}
\draw[->] (0,0) -- (0.5,0);
\draw[->] (0.5,0) -- (1.5,0);
\draw (1.5,0) -- (2,0);
\filldraw[white] (1,0) circle (3pt);
\draw (0.9,-0.1) rectangle ++(0.2,0.2);
\end{tikzpicture}=-\frac{ik^2}{2M}\,,
\end{equation}
\begin{equation}
\begin{tikzpicture}[baseline={([yshift=-.5ex]current bounding box.center)}]
\draw[->] (0,0) node [anchor=north] {$\mathbf{k}$} -- (0.5,0);
\draw[->] (0.5,0) -- (1.5,0);
\draw (1.5,0) -- (2,0) node [anchor=north] {$\mathbf{k}+\mathbf{q}-\mathbf{l}$};
\draw[gluon] (0,1) node [anchor=west] {$\,\,\,\mathbf{q},i$} -- (1,0);
\draw[gluon] (1,0) -- (2,1) node [anchor=east] {$\mathbf{l},j\,\,\,$};
\filldraw[white] (1,0) circle (3pt);
\draw (0.9,-0.1) rectangle ++(0.2,0.2);
\end{tikzpicture}=-\frac{ig^2\delta^{ij}}{2M}\,,
\end{equation}
\begin{equation}
\begin{tikzpicture}[baseline={([yshift=-.5ex]current bounding box.center)}]
\draw[->] (0,0) node [anchor=north] {$\mathbf{k}$} -- (0.5,0);
\draw[->] (0.5,0) -- (1.5,0);
\draw (1.5,0) -- (2,0) node [anchor=north] {$\mathbf{k}+\mathbf{q}$};
\draw[gluon] (1,1) node [anchor=west] {$\,\,\,\mathbf{q},i$} -- (1,0);
\filldraw[white] (1,0) circle (3pt);
\draw (0.9,-0.1) rectangle ++(0.2,0.2);
\end{tikzpicture}=-\frac{ig(2k^i+q^i)}{2M}\,,
\label{eq:a8}
\end{equation}
\begin{equation}
\begin{tikzpicture}
\draw[->] (0,0) -- (0.5,0);
\draw[->] (0.5,0) -- (1.5,0);
\draw (1.5,0) -- (2,0);
\filldraw[black] (1,0) circle (3pt);
\end{tikzpicture}=\frac{ic_4p^4}{8M^3}\,,
\end{equation}
and finally
\begin{equation}
\begin{tikzpicture}[baseline={([yshift=-.5ex]current bounding box.center)}]
\draw[->] (0,0) node [anchor=north] {$\mathbf{k}$} -- (0.5,0);
\draw[->] (0.5,0) -- (1.5,0);
\draw (1.5,0) -- (2,0) node [anchor=north] {$\mathbf{k}+\mathbf{q}$};
\draw[gluon] (1,1) node [anchor=west] {$\,\,\,\mathbf{q},i$} -- (1,0);
\filldraw[black] (1,0) circle (3pt);
\end{tikzpicture}=\frac{c_Fg(\mathbf{q}\times\boldsymbol{\sigma})}{2M}\,.
\end{equation}
\section{Gauge invariance of $\mathcal{I}$}
\label{sec:gauge}
In this appendix, we are going to discuss the gauge invariance of eq. (\ref{eq:I}) in more detail. Let us focus in the simplest term that is yet already problematic
\begin{equation}
    -i\Gamma\int\,d^4x\xi^\dagger_2\xi_1\,.
    \label{eq:Gammaterm}
\end{equation}
As we discussed, this terms is only invariant under a restricted class of gauge transformations, those that transform in the same way fields in the two branches of the Schwinger-Keldysh contour. This term can be made gauge-invariant by including two extra Wilson line.
\begin{equation}
    -i\Gamma\int\,d^4x\xi^\dagger_2 U_2(\mathbf{x};t,\infty)U_1(\mathbf{x};\infty,t)\xi_1\,,
\end{equation}
where $U_i$ is a temporal Wilson line made of fields of type $i$. This term is gauge-invariant because fields of type $1$ at $t\to\infty$ are identical to fields of type $2$ at $t\to\infty$. This is a consequence of the largest time equation \cite{Veltman:1963th} that implies that any correlator in which a \textit{quantum} field (in the sense of eq. (\ref{eq:defq})) has the largest time argument is zero. In other words, \textit{quantum} fields are zero at $t\to\infty$ and so are their gauge transformation.

Let us now discuss the differences with \cite{Crossley:2015evo} and other words regarding an EFT for hydrodynamics. There fields of type $1$ (or $2$) were assumed to appear always in gauge-invariant conbinations thanks to the introduction of Stueckelberg-like fields \cite{Ruegg:2003ps}. However, this solution is not suitable to our case since the generalization of the Stueckelberg theory to non-Abelian symmetries is problematic \cite{Ruegg:2003ps}. Note that in \cite{Crossley:2015evo} they are dealing with a theory for the field $A_\mu$ in which only derivatives of this field can appear in the influence functional. This is not analog to our case in which since the field $\xi$ does appear without derivates in most of the terms we are interested.

Now we can discuss the issue if the fields in $D_0$ and $\mathbf{B}$ appearing in eq. (\ref{eq:I}) are of type $1$ or $2$. Let us first note that if we impose that $S_{LET}=0$ in the limit $\xi_Q=A_Q=0$ then $\mathbf{B}$ and the $A_0$ field inside $D_0$ need to be evaluated at time $t$. Regarding $D_0$, there is only one possible combination because it can be checked that
\begin{equation}
    \xi^\dagger_2D_{0,2}U_2(\mathbf{x};t,\infty)U_1(\mathbf{x};\infty,t)\xi_1=\xi^\dagger_2 U_2(\mathbf{x};t,\infty)U_1(\mathbf{x};\infty,t)D_{0,1}\xi_1\,,
\end{equation}
where $D_{0,i}=\partial_0-igA_{0,1}$. Regarding $B$, it can enter in the following combinations
\begin{equation}
\begin{split}
    \xi^\dagger_2 B_2U_2(\mathbf{x};t,\infty)U_1(\mathbf{x};\infty,t)\xi_1\,, \\
    \xi^\dagger_2 U_2(\mathbf{x};t,\infty)U_1(\mathbf{x};\infty,t)B_1\xi_1\,.
\end{split}
\end{equation}
Therefore, to be more precised, in eq. (\ref{eq:I}) we should change
\begin{equation}
    \int\,d^4x\xi^\dagger_2\frac{\boldsymbol{\sigma}\cdot\mathbf{B}}{M}\xi_1\,,
\end{equation}
into
\begin{equation}
    \begin{split}
    \frac{1}{2}\int\,d^4x\xi^\dagger_2\frac{\boldsymbol{\sigma}\cdot\mathbf{B_2}}{M}U_2(\mathbf{x};t,\infty)U_1(\mathbf{x};\infty,t)\xi_1\\
    +\frac{1}{2}\int\,d^4x\xi^\dagger_2U_2(\mathbf{x};t,\infty)U_1(\mathbf{x};\infty,t)\frac{\boldsymbol{\sigma}\cdot\mathbf{B_1}}{M}\xi_1\,.
    \end{split}
\end{equation}

\bibliography{LangEFT.bib}

\begin{thebibliography}{55}%
\makeatletter
\providecommand \@ifxundefined [1]{%
 \@ifx{#1\undefined}
}%
\providecommand \@ifnum [1]{%
 \ifnum #1\expandafter \@firstoftwo
 \else \expandafter \@secondoftwo
 \fi
}%
\providecommand \@ifx [1]{%
 \ifx #1\expandafter \@firstoftwo
 \else \expandafter \@secondoftwo
 \fi
}%
\providecommand \natexlab [1]{#1}%
\providecommand \enquote  [1]{``#1''}%
\providecommand \bibnamefont  [1]{#1}%
\providecommand \bibfnamefont [1]{#1}%
\providecommand \citenamefont [1]{#1}%
\providecommand \href@noop [0]{\@secondoftwo}%
\providecommand \href [0]{\begingroup \@sanitize@url \@href}%
\providecommand \@href[1]{\@@startlink{#1}\@@href}%
\providecommand \@@href[1]{\endgroup#1\@@endlink}%
\providecommand \@sanitize@url [0]{\catcode `\\12\catcode `\$12\catcode
  `\&12\catcode `\#12\catcode `\^12\catcode `\_12\catcode `\%12\relax}%
\providecommand \@@startlink[1]{}%
\providecommand \@@endlink[0]{}%
\providecommand \url  [0]{\begingroup\@sanitize@url \@url }%
\providecommand \@url [1]{\endgroup\@href {#1}{\urlprefix }}%
\providecommand \urlprefix  [0]{URL }%
\providecommand \Eprint [0]{\href }%
\providecommand \doibase [0]{https://doi.org/}%
\providecommand \selectlanguage [0]{\@gobble}%
\providecommand \bibinfo  [0]{\@secondoftwo}%
\providecommand \bibfield  [0]{\@secondoftwo}%
\providecommand \translation [1]{[#1]}%
\providecommand \BibitemOpen [0]{}%
\providecommand \bibitemStop [0]{}%
\providecommand \bibitemNoStop [0]{.\EOS\space}%
\providecommand \EOS [0]{\spacefactor3000\relax}%
\providecommand \BibitemShut  [1]{\csname bibitem#1\endcsname}%
\let\auto@bib@innerbib\@empty
\bibitem [{\citenamefont {Weinberg}(1979)}]{Weinberg:1978kz}%
  \BibitemOpen
  \bibfield  {author} {\bibinfo {author} {\bibfnamefont {S.}~\bibnamefont
  {Weinberg}},\ }\bibfield  {title} {\bibinfo {title} {{Phenomenological
  Lagrangians}},\ }\href {https://doi.org/10.1016/0378-4371(79)90223-1}
  {\bibfield  {journal} {\bibinfo  {journal} {Physica A}\ }\textbf {\bibinfo
  {volume} {96}},\ \bibinfo {pages} {327} (\bibinfo {year} {1979})}\BibitemShut
  {NoStop}%
\bibitem [{\citenamefont {Caswell}\ and\ \citenamefont
  {Lepage}(1986)}]{Caswell:1985ui}%
  \BibitemOpen
  \bibfield  {author} {\bibinfo {author} {\bibfnamefont {W.~E.}\ \bibnamefont
  {Caswell}}\ and\ \bibinfo {author} {\bibfnamefont {G.~P.}\ \bibnamefont
  {Lepage}},\ }\bibfield  {title} {\bibinfo {title} {{Effective Lagrangians for
  Bound State Problems in QED, QCD, and Other Field Theories}},\ }\href
  {https://doi.org/10.1016/0370-2693(86)91297-9} {\bibfield  {journal}
  {\bibinfo  {journal} {Phys. Lett. B}\ }\textbf {\bibinfo {volume} {167}},\
  \bibinfo {pages} {437} (\bibinfo {year} {1986})}\BibitemShut {NoStop}%
\bibitem [{\citenamefont {Bodwin}\ \emph {et~al.}(1995)\citenamefont {Bodwin},
  \citenamefont {Braaten},\ and\ \citenamefont {Lepage}}]{Bodwin:1994jh}%
  \BibitemOpen
  \bibfield  {author} {\bibinfo {author} {\bibfnamefont {G.~T.}\ \bibnamefont
  {Bodwin}}, \bibinfo {author} {\bibfnamefont {E.}~\bibnamefont {Braaten}},\
  and\ \bibinfo {author} {\bibfnamefont {G.~P.}\ \bibnamefont {Lepage}},\
  }\bibfield  {title} {\bibinfo {title} {{Rigorous QCD analysis of inclusive
  annihilation and production of heavy quarkonium}},\ }\href
  {https://doi.org/10.1103/PhysRevD.55.5853} {\bibfield  {journal} {\bibinfo
  {journal} {Phys. Rev. D}\ }\textbf {\bibinfo {volume} {51}},\ \bibinfo
  {pages} {1125} (\bibinfo {year} {1995})},\ \bibinfo {note} {[Erratum:
  Phys.Rev.D 55, 5853 (1997)]},\ \Eprint {https://arxiv.org/abs/hep-ph/9407339}
  {arXiv:hep-ph/9407339} \BibitemShut {NoStop}%
\bibitem [{\citenamefont {Pineda}\ and\ \citenamefont
  {Soto}(1998)}]{Pineda:1997bj}%
  \BibitemOpen
  \bibfield  {author} {\bibinfo {author} {\bibfnamefont {A.}~\bibnamefont
  {Pineda}}\ and\ \bibinfo {author} {\bibfnamefont {J.}~\bibnamefont {Soto}},\
  }\bibfield  {title} {\bibinfo {title} {{Effective field theory for ultrasoft
  momenta in NRQCD and NRQED}},\ }\href
  {https://doi.org/10.1016/S0920-5632(97)01102-X} {\bibfield  {journal}
  {\bibinfo  {journal} {Nucl. Phys. B Proc. Suppl.}\ }\textbf {\bibinfo
  {volume} {64}},\ \bibinfo {pages} {428} (\bibinfo {year} {1998})},\ \Eprint
  {https://arxiv.org/abs/hep-ph/9707481} {arXiv:hep-ph/9707481} \BibitemShut
  {NoStop}%
\bibitem [{\citenamefont {Brambilla}\ \emph {et~al.}(2000)\citenamefont
  {Brambilla}, \citenamefont {Pineda}, \citenamefont {Soto},\ and\
  \citenamefont {Vairo}}]{Brambilla:1999xf}%
  \BibitemOpen
  \bibfield  {author} {\bibinfo {author} {\bibfnamefont {N.}~\bibnamefont
  {Brambilla}}, \bibinfo {author} {\bibfnamefont {A.}~\bibnamefont {Pineda}},
  \bibinfo {author} {\bibfnamefont {J.}~\bibnamefont {Soto}},\ and\ \bibinfo
  {author} {\bibfnamefont {A.}~\bibnamefont {Vairo}},\ }\bibfield  {title}
  {\bibinfo {title} {{Potential NRQCD: An Effective theory for heavy
  quarkonium}},\ }\href {https://doi.org/10.1016/S0550-3213(99)00693-8}
  {\bibfield  {journal} {\bibinfo  {journal} {Nucl. Phys. B}\ }\textbf
  {\bibinfo {volume} {566}},\ \bibinfo {pages} {275} (\bibinfo {year}
  {2000})},\ \Eprint {https://arxiv.org/abs/hep-ph/9907240}
  {arXiv:hep-ph/9907240} \BibitemShut {NoStop}%
\bibitem [{\citenamefont {Brambilla}\ \emph {et~al.}(2005)\citenamefont
  {Brambilla}, \citenamefont {Pineda}, \citenamefont {Soto},\ and\
  \citenamefont {Vairo}}]{Brambilla:2004jw}%
  \BibitemOpen
  \bibfield  {author} {\bibinfo {author} {\bibfnamefont {N.}~\bibnamefont
  {Brambilla}}, \bibinfo {author} {\bibfnamefont {A.}~\bibnamefont {Pineda}},
  \bibinfo {author} {\bibfnamefont {J.}~\bibnamefont {Soto}},\ and\ \bibinfo
  {author} {\bibfnamefont {A.}~\bibnamefont {Vairo}},\ }\bibfield  {title}
  {\bibinfo {title} {{Effective Field Theories for Heavy Quarkonium}},\ }\href
  {https://doi.org/10.1103/RevModPhys.77.1423} {\bibfield  {journal} {\bibinfo
  {journal} {Rev. Mod. Phys.}\ }\textbf {\bibinfo {volume} {77}},\ \bibinfo
  {pages} {1423} (\bibinfo {year} {2005})},\ \Eprint
  {https://arxiv.org/abs/hep-ph/0410047} {arXiv:hep-ph/0410047} \BibitemShut
  {NoStop}%
\bibitem [{\citenamefont {Escobedo}\ and\ \citenamefont
  {Soto}(2008)}]{Escobedo:2008sy}%
  \BibitemOpen
  \bibfield  {author} {\bibinfo {author} {\bibfnamefont {M.~A.}\ \bibnamefont
  {Escobedo}}\ and\ \bibinfo {author} {\bibfnamefont {J.}~\bibnamefont
  {Soto}},\ }\bibfield  {title} {\bibinfo {title} {{Non-relativistic bound
  states at finite temperature (I): The Hydrogen atom}},\ }\href
  {https://doi.org/10.1103/PhysRevA.78.032520} {\bibfield  {journal} {\bibinfo
  {journal} {Phys. Rev. A}\ }\textbf {\bibinfo {volume} {78}},\ \bibinfo
  {pages} {032520} (\bibinfo {year} {2008})},\ \Eprint
  {https://arxiv.org/abs/0804.0691} {arXiv:0804.0691 [hep-ph]} \BibitemShut
  {NoStop}%
\bibitem [{\citenamefont {Brambilla}\ \emph {et~al.}(2008)\citenamefont
  {Brambilla}, \citenamefont {Ghiglieri}, \citenamefont {Vairo},\ and\
  \citenamefont {Petreczky}}]{Brambilla:2008cx}%
  \BibitemOpen
  \bibfield  {author} {\bibinfo {author} {\bibfnamefont {N.}~\bibnamefont
  {Brambilla}}, \bibinfo {author} {\bibfnamefont {J.}~\bibnamefont
  {Ghiglieri}}, \bibinfo {author} {\bibfnamefont {A.}~\bibnamefont {Vairo}},\
  and\ \bibinfo {author} {\bibfnamefont {P.}~\bibnamefont {Petreczky}},\
  }\bibfield  {title} {\bibinfo {title} {{Static quark-antiquark pairs at
  finite temperature}},\ }\href {https://doi.org/10.1103/PhysRevD.78.014017}
  {\bibfield  {journal} {\bibinfo  {journal} {Phys. Rev. D}\ }\textbf {\bibinfo
  {volume} {78}},\ \bibinfo {pages} {014017} (\bibinfo {year} {2008})},\
  \Eprint {https://arxiv.org/abs/0804.0993} {arXiv:0804.0993 [hep-ph]}
  \BibitemShut {NoStop}%
\bibitem [{\citenamefont {Brambilla}\ \emph {et~al.}(2010)\citenamefont
  {Brambilla}, \citenamefont {Escobedo}, \citenamefont {Ghiglieri},
  \citenamefont {Soto},\ and\ \citenamefont {Vairo}}]{Brambilla:2010vq}%
  \BibitemOpen
  \bibfield  {author} {\bibinfo {author} {\bibfnamefont {N.}~\bibnamefont
  {Brambilla}}, \bibinfo {author} {\bibfnamefont {M.~A.}\ \bibnamefont
  {Escobedo}}, \bibinfo {author} {\bibfnamefont {J.}~\bibnamefont {Ghiglieri}},
  \bibinfo {author} {\bibfnamefont {J.}~\bibnamefont {Soto}},\ and\ \bibinfo
  {author} {\bibfnamefont {A.}~\bibnamefont {Vairo}},\ }\bibfield  {title}
  {\bibinfo {title} {{Heavy Quarkonium in a weakly-coupled quark-gluon plasma
  below the melting temperature}},\ }\href
  {https://doi.org/10.1007/JHEP09(2010)038} {\bibfield  {journal} {\bibinfo
  {journal} {JHEP}\ }\textbf {\bibinfo {volume} {09}},\ \bibinfo {pages}
  {038}},\ \Eprint {https://arxiv.org/abs/1007.4156} {arXiv:1007.4156 [hep-ph]}
  \BibitemShut {NoStop}%
\bibitem [{\citenamefont {Brambilla}\ \emph {et~al.}(2017)\citenamefont
  {Brambilla}, \citenamefont {Escobedo}, \citenamefont {Soto},\ and\
  \citenamefont {Vairo}}]{Brambilla:2016wgg}%
  \BibitemOpen
  \bibfield  {author} {\bibinfo {author} {\bibfnamefont {N.}~\bibnamefont
  {Brambilla}}, \bibinfo {author} {\bibfnamefont {M.~A.}\ \bibnamefont
  {Escobedo}}, \bibinfo {author} {\bibfnamefont {J.}~\bibnamefont {Soto}},\
  and\ \bibinfo {author} {\bibfnamefont {A.}~\bibnamefont {Vairo}},\ }\bibfield
   {title} {\bibinfo {title} {{Quarkonium suppression in heavy-ion collisions:
  an open quantum system approach}},\ }\href
  {https://doi.org/10.1103/PhysRevD.96.034021} {\bibfield  {journal} {\bibinfo
  {journal} {Phys. Rev. D}\ }\textbf {\bibinfo {volume} {96}},\ \bibinfo
  {pages} {034021} (\bibinfo {year} {2017})},\ \Eprint
  {https://arxiv.org/abs/1612.07248} {arXiv:1612.07248 [hep-ph]} \BibitemShut
  {NoStop}%
\bibitem [{\citenamefont {Brambilla}\ \emph {et~al.}(2018)\citenamefont
  {Brambilla}, \citenamefont {Escobedo}, \citenamefont {Soto},\ and\
  \citenamefont {Vairo}}]{Brambilla:2017zei}%
  \BibitemOpen
  \bibfield  {author} {\bibinfo {author} {\bibfnamefont {N.}~\bibnamefont
  {Brambilla}}, \bibinfo {author} {\bibfnamefont {M.~A.}\ \bibnamefont
  {Escobedo}}, \bibinfo {author} {\bibfnamefont {J.}~\bibnamefont {Soto}},\
  and\ \bibinfo {author} {\bibfnamefont {A.}~\bibnamefont {Vairo}},\ }\bibfield
   {title} {\bibinfo {title} {{Heavy quarkonium suppression in a fireball}},\
  }\href {https://doi.org/10.1103/PhysRevD.97.074009} {\bibfield  {journal}
  {\bibinfo  {journal} {Phys. Rev. D}\ }\textbf {\bibinfo {volume} {97}},\
  \bibinfo {pages} {074009} (\bibinfo {year} {2018})},\ \Eprint
  {https://arxiv.org/abs/1711.04515} {arXiv:1711.04515 [hep-ph]} \BibitemShut
  {NoStop}%
\bibitem [{\citenamefont {Yao}\ and\ \citenamefont
  {M\"uller}(2018)}]{Yao:2017fuc}%
  \BibitemOpen
  \bibfield  {author} {\bibinfo {author} {\bibfnamefont {X.}~\bibnamefont
  {Yao}}\ and\ \bibinfo {author} {\bibfnamefont {B.}~\bibnamefont {M\"uller}},\
  }\bibfield  {title} {\bibinfo {title} {{Approach to equilibrium of quarkonium
  in quark-gluon plasma}},\ }\href {https://doi.org/10.1103/PhysRevC.97.014908}
  {\bibfield  {journal} {\bibinfo  {journal} {Phys. Rev. C}\ }\textbf {\bibinfo
  {volume} {97}},\ \bibinfo {pages} {014908} (\bibinfo {year} {2018})},\
  \bibinfo {note} {[Erratum: Phys.Rev.C 97, 049903 (2018)]},\ \Eprint
  {https://arxiv.org/abs/1709.03529} {arXiv:1709.03529 [hep-ph]} \BibitemShut
  {NoStop}%
\bibitem [{\citenamefont {Yao}\ and\ \citenamefont
  {Mehen}(2019)}]{Yao:2018nmy}%
  \BibitemOpen
  \bibfield  {author} {\bibinfo {author} {\bibfnamefont {X.}~\bibnamefont
  {Yao}}\ and\ \bibinfo {author} {\bibfnamefont {T.}~\bibnamefont {Mehen}},\
  }\bibfield  {title} {\bibinfo {title} {{Quarkonium in-medium transport
  equation derived from first principles}},\ }\href
  {https://doi.org/10.1103/PhysRevD.99.096028} {\bibfield  {journal} {\bibinfo
  {journal} {Phys. Rev. D}\ }\textbf {\bibinfo {volume} {99}},\ \bibinfo
  {pages} {096028} (\bibinfo {year} {2019})},\ \Eprint
  {https://arxiv.org/abs/1811.07027} {arXiv:1811.07027 [hep-ph]} \BibitemShut
  {NoStop}%
\bibitem [{\citenamefont {Vaidya}\ and\ \citenamefont
  {Yao}(2020)}]{Vaidya:2020cyi}%
  \BibitemOpen
  \bibfield  {author} {\bibinfo {author} {\bibfnamefont {V.}~\bibnamefont
  {Vaidya}}\ and\ \bibinfo {author} {\bibfnamefont {X.}~\bibnamefont {Yao}},\
  }\bibfield  {title} {\bibinfo {title} {{Transverse momentum broadening of a
  jet in quark-gluon plasma: an open quantum system EFT}},\ }\href
  {https://doi.org/10.1007/JHEP10(2020)024} {\bibfield  {journal} {\bibinfo
  {journal} {JHEP}\ }\textbf {\bibinfo {volume} {10}},\ \bibinfo {pages}
  {024}},\ \Eprint {https://arxiv.org/abs/2004.11403} {arXiv:2004.11403
  [hep-ph]} \BibitemShut {NoStop}%
\bibitem [{\citenamefont {Vaidya}(2021)}]{Vaidya:2020lih}%
  \BibitemOpen
  \bibfield  {author} {\bibinfo {author} {\bibfnamefont {V.}~\bibnamefont
  {Vaidya}},\ }\bibfield  {title} {\bibinfo {title} {{Effective Field Theory
  for jet substructure in heavy ion collisions}},\ }\href
  {https://doi.org/10.1007/JHEP11(2021)064} {\bibfield  {journal} {\bibinfo
  {journal} {JHEP}\ }\textbf {\bibinfo {volume} {11}},\ \bibinfo {pages}
  {064}},\ \Eprint {https://arxiv.org/abs/2010.00028} {arXiv:2010.00028
  [hep-ph]} \BibitemShut {NoStop}%
\bibitem [{\citenamefont {Bellac}(2011)}]{Bellac:2011kqa}%
  \BibitemOpen
  \bibfield  {author} {\bibinfo {author} {\bibfnamefont {M.~L.}\ \bibnamefont
  {Bellac}},\ }\href {https://doi.org/10.1017/CBO9780511721700} {\emph
  {\bibinfo {title} {{Thermal Field Theory}}}},\ Cambridge Monographs on
  Mathematical Physics\ (\bibinfo  {publisher} {Cambridge University Press},\
  \bibinfo {year} {2011})\BibitemShut {NoStop}%
\bibitem [{\citenamefont {Breuer}\ and\ \citenamefont
  {Petruccione}(2007)}]{Breuer:2007juk}%
  \BibitemOpen
  \bibfield  {author} {\bibinfo {author} {\bibfnamefont {H.-P.}\ \bibnamefont
  {Breuer}}\ and\ \bibinfo {author} {\bibfnamefont {F.}~\bibnamefont
  {Petruccione}},\ }\href
  {https://doi.org/10.1093/acprof:oso/9780199213900.001.0001} {\emph {\bibinfo
  {title} {{The Theory of Open Quantum Systems}}}}\ (\bibinfo  {publisher}
  {Oxford University Press},\ \bibinfo {year} {2007})\BibitemShut {NoStop}%
\bibitem [{\citenamefont {Feynman}\ and\ \citenamefont
  {Vernon}(1963)}]{Feynman:1963fq}%
  \BibitemOpen
  \bibfield  {author} {\bibinfo {author} {\bibfnamefont {R.~P.}\ \bibnamefont
  {Feynman}}\ and\ \bibinfo {author} {\bibfnamefont {F.~L.}\ \bibnamefont
  {Vernon}, \bibfnamefont {Jr.}},\ }\bibfield  {title} {\bibinfo {title} {{The
  Theory of a general quantum system interacting with a linear dissipative
  system}},\ }\href {https://doi.org/10.1016/0003-4916(63)90068-X} {\bibfield
  {journal} {\bibinfo  {journal} {Annals Phys.}\ }\textbf {\bibinfo {volume}
  {24}},\ \bibinfo {pages} {118} (\bibinfo {year} {1963})}\BibitemShut
  {NoStop}%
\bibitem [{\citenamefont {Baidya}\ \emph {et~al.}(2017)\citenamefont {Baidya},
  \citenamefont {Jana}, \citenamefont {Loganayagam},\ and\ \citenamefont
  {Rudra}}]{Baidya:2017eho}%
  \BibitemOpen
  \bibfield  {author} {\bibinfo {author} {\bibfnamefont {A.}~\bibnamefont
  {Baidya}}, \bibinfo {author} {\bibfnamefont {C.}~\bibnamefont {Jana}},
  \bibinfo {author} {\bibfnamefont {R.}~\bibnamefont {Loganayagam}},\ and\
  \bibinfo {author} {\bibfnamefont {A.}~\bibnamefont {Rudra}},\ }\bibfield
  {title} {\bibinfo {title} {{Renormalization in open quantum field theory.
  Part I. Scalar field theory}},\ }\href
  {https://doi.org/10.1007/JHEP11(2017)204} {\bibfield  {journal} {\bibinfo
  {journal} {JHEP}\ }\textbf {\bibinfo {volume} {11}},\ \bibinfo {pages}
  {204}},\ \Eprint {https://arxiv.org/abs/1704.08335} {arXiv:1704.08335
  [hep-th]} \BibitemShut {NoStop}%
\bibitem [{\citenamefont {Haehl}\ \emph {et~al.}(2017)\citenamefont {Haehl},
  \citenamefont {Loganayagam},\ and\ \citenamefont
  {Rangamani}}]{Haehl:2016pec}%
  \BibitemOpen
  \bibfield  {author} {\bibinfo {author} {\bibfnamefont {F.~M.}\ \bibnamefont
  {Haehl}}, \bibinfo {author} {\bibfnamefont {R.}~\bibnamefont {Loganayagam}},\
  and\ \bibinfo {author} {\bibfnamefont {M.}~\bibnamefont {Rangamani}},\
  }\bibfield  {title} {\bibinfo {title} {{Schwinger-Keldysh formalism. Part I:
  BRST symmetries and superspace}},\ }\href
  {https://doi.org/10.1007/JHEP06(2017)069} {\bibfield  {journal} {\bibinfo
  {journal} {JHEP}\ }\textbf {\bibinfo {volume} {06}},\ \bibinfo {pages}
  {069}},\ \Eprint {https://arxiv.org/abs/1610.01940} {arXiv:1610.01940
  [hep-th]} \BibitemShut {NoStop}%
\bibitem [{\citenamefont {Crossley}\ \emph {et~al.}(2017)\citenamefont
  {Crossley}, \citenamefont {Glorioso},\ and\ \citenamefont
  {Liu}}]{Crossley:2015evo}%
  \BibitemOpen
  \bibfield  {author} {\bibinfo {author} {\bibfnamefont {M.}~\bibnamefont
  {Crossley}}, \bibinfo {author} {\bibfnamefont {P.}~\bibnamefont {Glorioso}},\
  and\ \bibinfo {author} {\bibfnamefont {H.}~\bibnamefont {Liu}},\ }\bibfield
  {title} {\bibinfo {title} {{Effective field theory of dissipative fluids}},\
  }\href {https://doi.org/10.1007/JHEP09(2017)095} {\bibfield  {journal}
  {\bibinfo  {journal} {JHEP}\ }\textbf {\bibinfo {volume} {09}},\ \bibinfo
  {pages} {095}},\ \Eprint {https://arxiv.org/abs/1511.03646} {arXiv:1511.03646
  [hep-th]} \BibitemShut {NoStop}%
\bibitem [{\citenamefont {Jensen}\ \emph {et~al.}(2018)\citenamefont {Jensen},
  \citenamefont {Pinzani-Fokeeva},\ and\ \citenamefont
  {Yarom}}]{Jensen:2017kzi}%
  \BibitemOpen
  \bibfield  {author} {\bibinfo {author} {\bibfnamefont {K.}~\bibnamefont
  {Jensen}}, \bibinfo {author} {\bibfnamefont {N.}~\bibnamefont
  {Pinzani-Fokeeva}},\ and\ \bibinfo {author} {\bibfnamefont {A.}~\bibnamefont
  {Yarom}},\ }\bibfield  {title} {\bibinfo {title} {{Dissipative hydrodynamics
  in superspace}},\ }\href {https://doi.org/10.1007/JHEP09(2018)127} {\bibfield
   {journal} {\bibinfo  {journal} {JHEP}\ }\textbf {\bibinfo {volume} {09}},\
  \bibinfo {pages} {127}},\ \Eprint {https://arxiv.org/abs/1701.07436}
  {arXiv:1701.07436 [hep-th]} \BibitemShut {NoStop}%
\bibitem [{\citenamefont {Svetitsky}(1988)}]{Svetitsky:1987gq}%
  \BibitemOpen
  \bibfield  {author} {\bibinfo {author} {\bibfnamefont {B.}~\bibnamefont
  {Svetitsky}},\ }\bibfield  {title} {\bibinfo {title} {{Diffusion of charmed
  quarks in the quark-gluon plasma}},\ }\href
  {https://doi.org/10.1103/PhysRevD.37.2484} {\bibfield  {journal} {\bibinfo
  {journal} {Phys. Rev. D}\ }\textbf {\bibinfo {volume} {37}},\ \bibinfo
  {pages} {2484} (\bibinfo {year} {1988})}\BibitemShut {NoStop}%
\bibitem [{\citenamefont {Torres-Rincon}(2012)}]{Torres-Rincon:2012sda}%
  \BibitemOpen
  \bibfield  {author} {\bibinfo {author} {\bibfnamefont {J.~M.}\ \bibnamefont
  {Torres-Rincon}},\ }\emph {\bibinfo {title} {{Hadronic transport coefficients
  from effective field theories}}},\ \href
  {https://doi.org/10.1007/978-3-319-00425-9} {Ph.D. thesis} (\bibinfo {year}
  {2012}),\ \Eprint {https://arxiv.org/abs/1205.0782} {arXiv:1205.0782
  [hep-ph]} \BibitemShut {NoStop}%
\bibitem [{\citenamefont {Moore}\ and\ \citenamefont
  {Teaney}(2005)}]{Moore:2004tg}%
  \BibitemOpen
  \bibfield  {author} {\bibinfo {author} {\bibfnamefont {G.~D.}\ \bibnamefont
  {Moore}}\ and\ \bibinfo {author} {\bibfnamefont {D.}~\bibnamefont {Teaney}},\
  }\bibfield  {title} {\bibinfo {title} {{How much do heavy quarks thermalize
  in a heavy ion collision?}},\ }\href
  {https://doi.org/10.1103/PhysRevC.71.064904} {\bibfield  {journal} {\bibinfo
  {journal} {Phys. Rev. C}\ }\textbf {\bibinfo {volume} {71}},\ \bibinfo
  {pages} {064904} (\bibinfo {year} {2005})},\ \Eprint
  {https://arxiv.org/abs/hep-ph/0412346} {arXiv:hep-ph/0412346} \BibitemShut
  {NoStop}%
\bibitem [{\citenamefont {van Hees}\ \emph {et~al.}(2006)\citenamefont {van
  Hees}, \citenamefont {Greco},\ and\ \citenamefont {Rapp}}]{vanHees:2005wb}%
  \BibitemOpen
  \bibfield  {author} {\bibinfo {author} {\bibfnamefont {H.}~\bibnamefont {van
  Hees}}, \bibinfo {author} {\bibfnamefont {V.}~\bibnamefont {Greco}},\ and\
  \bibinfo {author} {\bibfnamefont {R.}~\bibnamefont {Rapp}},\ }\bibfield
  {title} {\bibinfo {title} {{Heavy-quark probes of the quark-gluon plasma at
  RHIC}},\ }\href {https://doi.org/10.1103/PhysRevC.73.034913} {\bibfield
  {journal} {\bibinfo  {journal} {Phys. Rev. C}\ }\textbf {\bibinfo {volume}
  {73}},\ \bibinfo {pages} {034913} (\bibinfo {year} {2006})},\ \Eprint
  {https://arxiv.org/abs/nucl-th/0508055} {arXiv:nucl-th/0508055} \BibitemShut
  {NoStop}%
\bibitem [{\citenamefont {Rapp}\ and\ \citenamefont {van
  Hees}(2008)}]{Rapp:2008qc}%
  \BibitemOpen
  \bibfield  {author} {\bibinfo {author} {\bibfnamefont {R.}~\bibnamefont
  {Rapp}}\ and\ \bibinfo {author} {\bibfnamefont {H.}~\bibnamefont {van
  Hees}},\ }\bibfield  {title} {\bibinfo {title} {{Heavy Quark Diffusion as a
  Probe of the Quark-Gluon Plasma}},\ }\href@noop {} {\  (\bibinfo {year}
  {2008})},\ \Eprint {https://arxiv.org/abs/0803.0901} {arXiv:0803.0901
  [hep-ph]} \BibitemShut {NoStop}%
\bibitem [{\citenamefont {Akamatsu}\ \emph {et~al.}(2009)\citenamefont
  {Akamatsu}, \citenamefont {Hatsuda},\ and\ \citenamefont
  {Hirano}}]{Akamatsu:2008ge}%
  \BibitemOpen
  \bibfield  {author} {\bibinfo {author} {\bibfnamefont {Y.}~\bibnamefont
  {Akamatsu}}, \bibinfo {author} {\bibfnamefont {T.}~\bibnamefont {Hatsuda}},\
  and\ \bibinfo {author} {\bibfnamefont {T.}~\bibnamefont {Hirano}},\
  }\bibfield  {title} {\bibinfo {title} {{Heavy Quark Diffusion with
  Relativistic Langevin Dynamics in the Quark-Gluon Fluid}},\ }\href
  {https://doi.org/10.1103/PhysRevC.79.054907} {\bibfield  {journal} {\bibinfo
  {journal} {Phys. Rev. C}\ }\textbf {\bibinfo {volume} {79}},\ \bibinfo
  {pages} {054907} (\bibinfo {year} {2009})},\ \Eprint
  {https://arxiv.org/abs/0809.1499} {arXiv:0809.1499 [hep-ph]} \BibitemShut
  {NoStop}%
\bibitem [{\citenamefont {Young}\ and\ \citenamefont
  {Shuryak}(2009)}]{Young:2008he}%
  \BibitemOpen
  \bibfield  {author} {\bibinfo {author} {\bibfnamefont {C.}~\bibnamefont
  {Young}}\ and\ \bibinfo {author} {\bibfnamefont {E.}~\bibnamefont
  {Shuryak}},\ }\bibfield  {title} {\bibinfo {title} {{Charmonium in strongly
  coupled quark-gluon plasma}},\ }\href
  {https://doi.org/10.1103/PhysRevC.79.034907} {\bibfield  {journal} {\bibinfo
  {journal} {Phys. Rev. C}\ }\textbf {\bibinfo {volume} {79}},\ \bibinfo
  {pages} {034907} (\bibinfo {year} {2009})},\ \Eprint
  {https://arxiv.org/abs/0803.2866} {arXiv:0803.2866 [nucl-th]} \BibitemShut
  {NoStop}%
\bibitem [{\citenamefont {Bauer}\ \emph {et~al.}(2001)\citenamefont {Bauer},
  \citenamefont {Fleming}, \citenamefont {Pirjol},\ and\ \citenamefont
  {Stewart}}]{Bauer:2000yr}%
  \BibitemOpen
  \bibfield  {author} {\bibinfo {author} {\bibfnamefont {C.~W.}\ \bibnamefont
  {Bauer}}, \bibinfo {author} {\bibfnamefont {S.}~\bibnamefont {Fleming}},
  \bibinfo {author} {\bibfnamefont {D.}~\bibnamefont {Pirjol}},\ and\ \bibinfo
  {author} {\bibfnamefont {I.~W.}\ \bibnamefont {Stewart}},\ }\bibfield
  {title} {\bibinfo {title} {{An Effective field theory for collinear and soft
  gluons: Heavy to light decays}},\ }\href
  {https://doi.org/10.1103/PhysRevD.63.114020} {\bibfield  {journal} {\bibinfo
  {journal} {Phys. Rev. D}\ }\textbf {\bibinfo {volume} {63}},\ \bibinfo
  {pages} {114020} (\bibinfo {year} {2001})},\ \Eprint
  {https://arxiv.org/abs/hep-ph/0011336} {arXiv:hep-ph/0011336} \BibitemShut
  {NoStop}%
\bibitem [{\citenamefont {Bauer}\ \emph {et~al.}(2002)\citenamefont {Bauer},
  \citenamefont {Pirjol},\ and\ \citenamefont {Stewart}}]{Bauer:2001yt}%
  \BibitemOpen
  \bibfield  {author} {\bibinfo {author} {\bibfnamefont {C.~W.}\ \bibnamefont
  {Bauer}}, \bibinfo {author} {\bibfnamefont {D.}~\bibnamefont {Pirjol}},\ and\
  \bibinfo {author} {\bibfnamefont {I.~W.}\ \bibnamefont {Stewart}},\
  }\bibfield  {title} {\bibinfo {title} {{Soft collinear factorization in
  effective field theory}},\ }\href
  {https://doi.org/10.1103/PhysRevD.65.054022} {\bibfield  {journal} {\bibinfo
  {journal} {Phys. Rev. D}\ }\textbf {\bibinfo {volume} {65}},\ \bibinfo
  {pages} {054022} (\bibinfo {year} {2002})},\ \Eprint
  {https://arxiv.org/abs/hep-ph/0109045} {arXiv:hep-ph/0109045} \BibitemShut
  {NoStop}%
\bibitem [{\citenamefont {Manuel}\ and\ \citenamefont
  {Torres-Rincon}(2014)}]{Manuel:2014dza}%
  \BibitemOpen
  \bibfield  {author} {\bibinfo {author} {\bibfnamefont {C.}~\bibnamefont
  {Manuel}}\ and\ \bibinfo {author} {\bibfnamefont {J.~M.}\ \bibnamefont
  {Torres-Rincon}},\ }\bibfield  {title} {\bibinfo {title} {{Chiral transport
  equation from the quantum Dirac Hamiltonian and the on-shell effective field
  theory}},\ }\href {https://doi.org/10.1103/PhysRevD.90.076007} {\bibfield
  {journal} {\bibinfo  {journal} {Phys. Rev. D}\ }\textbf {\bibinfo {volume}
  {90}},\ \bibinfo {pages} {076007} (\bibinfo {year} {2014})},\ \Eprint
  {https://arxiv.org/abs/1404.6409} {arXiv:1404.6409 [hep-ph]} \BibitemShut
  {NoStop}%
\bibitem [{\citenamefont {Manuel}\ \emph {et~al.}(2016)\citenamefont {Manuel},
  \citenamefont {Soto},\ and\ \citenamefont {Stetina}}]{Manuel:2016wqs}%
  \BibitemOpen
  \bibfield  {author} {\bibinfo {author} {\bibfnamefont {C.}~\bibnamefont
  {Manuel}}, \bibinfo {author} {\bibfnamefont {J.}~\bibnamefont {Soto}},\ and\
  \bibinfo {author} {\bibfnamefont {S.}~\bibnamefont {Stetina}},\ }\bibfield
  {title} {\bibinfo {title} {{On-shell effective field theory: A systematic
  tool to compute power corrections to the hard thermal loops}},\ }\href
  {https://doi.org/10.1103/PhysRevD.94.025017} {\bibfield  {journal} {\bibinfo
  {journal} {Phys. Rev. D}\ }\textbf {\bibinfo {volume} {94}},\ \bibinfo
  {pages} {025017} (\bibinfo {year} {2016})},\ \bibinfo {note} {[Erratum:
  Phys.Rev.D 96, 129901 (2017)]},\ \Eprint {https://arxiv.org/abs/1603.05514}
  {arXiv:1603.05514 [hep-ph]} \BibitemShut {NoStop}%
\bibitem [{\citenamefont {Brambilla}\ \emph {et~al.}(2004)\citenamefont
  {Brambilla}, \citenamefont {Pineda}, \citenamefont {Soto},\ and\
  \citenamefont {Vairo}}]{Brambilla:2003mu}%
  \BibitemOpen
  \bibfield  {author} {\bibinfo {author} {\bibfnamefont {N.}~\bibnamefont
  {Brambilla}}, \bibinfo {author} {\bibfnamefont {A.}~\bibnamefont {Pineda}},
  \bibinfo {author} {\bibfnamefont {J.}~\bibnamefont {Soto}},\ and\ \bibinfo
  {author} {\bibfnamefont {A.}~\bibnamefont {Vairo}},\ }\bibfield  {title}
  {\bibinfo {title} {{The (m Lambda QCD)**1/2 scale in heavy quarkonium}},\
  }\href {https://doi.org/10.1016/j.physletb.2003.11.031} {\bibfield  {journal}
  {\bibinfo  {journal} {Phys. Lett. B}\ }\textbf {\bibinfo {volume} {580}},\
  \bibinfo {pages} {60} (\bibinfo {year} {2004})},\ \Eprint
  {https://arxiv.org/abs/hep-ph/0307159} {arXiv:hep-ph/0307159} \BibitemShut
  {NoStop}%
\bibitem [{\citenamefont {Escobedo}(2021)}]{Escobedo:2020tuc}%
  \BibitemOpen
  \bibfield  {author} {\bibinfo {author} {\bibfnamefont {M.~A.}\ \bibnamefont
  {Escobedo}},\ }\bibfield  {title} {\bibinfo {title} {{Medium evolution of a
  static quark-antiquark pair in the large $N_c$ limit}},\ }\href
  {https://doi.org/10.1103/PhysRevD.103.034010} {\bibfield  {journal} {\bibinfo
   {journal} {Phys. Rev. D}\ }\textbf {\bibinfo {volume} {103}},\ \bibinfo
  {pages} {034010} (\bibinfo {year} {2021})},\ \Eprint
  {https://arxiv.org/abs/2010.10424} {arXiv:2010.10424 [hep-ph]} \BibitemShut
  {NoStop}%
\bibitem [{\citenamefont {Bu}\ and\ \citenamefont {Zhang}(2021)}]{Bu:2021jlp}%
  \BibitemOpen
  \bibfield  {author} {\bibinfo {author} {\bibfnamefont {Y.}~\bibnamefont
  {Bu}}\ and\ \bibinfo {author} {\bibfnamefont {B.}~\bibnamefont {Zhang}},\
  }\bibfield  {title} {\bibinfo {title} {{Schwinger-Keldysh effective action
  for a relativistic Brownian particle in the AdS/CFT correspondence}},\ }\href
  {https://doi.org/10.1103/PhysRevD.104.086002} {\bibfield  {journal} {\bibinfo
   {journal} {Phys. Rev. D}\ }\textbf {\bibinfo {volume} {104}},\ \bibinfo
  {pages} {086002} (\bibinfo {year} {2021})},\ \Eprint
  {https://arxiv.org/abs/2108.10060} {arXiv:2108.10060 [hep-th]} \BibitemShut
  {NoStop}%
\bibitem [{\citenamefont {Luke}\ and\ \citenamefont
  {Manohar}(1992)}]{Luke:1992cs}%
  \BibitemOpen
  \bibfield  {author} {\bibinfo {author} {\bibfnamefont {M.~E.}\ \bibnamefont
  {Luke}}\ and\ \bibinfo {author} {\bibfnamefont {A.~V.}\ \bibnamefont
  {Manohar}},\ }\bibfield  {title} {\bibinfo {title} {{Reparametrization
  invariance constraints on heavy particle effective field theories}},\ }\href
  {https://doi.org/10.1016/0370-2693(92)91786-9} {\bibfield  {journal}
  {\bibinfo  {journal} {Phys. Lett. B}\ }\textbf {\bibinfo {volume} {286}},\
  \bibinfo {pages} {348} (\bibinfo {year} {1992})},\ \Eprint
  {https://arxiv.org/abs/hep-ph/9205228} {arXiv:hep-ph/9205228} \BibitemShut
  {NoStop}%
\bibitem [{\citenamefont {Vasak}\ \emph {et~al.}(1987)\citenamefont {Vasak},
  \citenamefont {Gyulassy},\ and\ \citenamefont {Elze}}]{Vasak:1987um}%
  \BibitemOpen
  \bibfield  {author} {\bibinfo {author} {\bibfnamefont {D.}~\bibnamefont
  {Vasak}}, \bibinfo {author} {\bibfnamefont {M.}~\bibnamefont {Gyulassy}},\
  and\ \bibinfo {author} {\bibfnamefont {H.~T.}\ \bibnamefont {Elze}},\
  }\bibfield  {title} {\bibinfo {title} {{Quantum Transport Theory for Abelian
  Plasmas}},\ }\href {https://doi.org/10.1016/0003-4916(87)90169-2} {\bibfield
  {journal} {\bibinfo  {journal} {Annals Phys.}\ }\textbf {\bibinfo {volume}
  {173}},\ \bibinfo {pages} {462} (\bibinfo {year} {1987})}\BibitemShut
  {NoStop}%
\bibitem [{\citenamefont {Keldysh}(1964)}]{Keldysh:1964ud}%
  \BibitemOpen
  \bibfield  {author} {\bibinfo {author} {\bibfnamefont {L.~V.}\ \bibnamefont
  {Keldysh}},\ }\bibfield  {title} {\bibinfo {title} {{Diagram technique for
  nonequilibrium processes}},\ }\href@noop {} {\bibfield  {journal} {\bibinfo
  {journal} {Zh. Eksp. Teor. Fiz.}\ }\textbf {\bibinfo {volume} {47}},\
  \bibinfo {pages} {1515} (\bibinfo {year} {1964})}\BibitemShut {NoStop}%
\bibitem [{\citenamefont {Braaten}\ and\ \citenamefont
  {Pisarski}(1992)}]{Braaten:1991gm}%
  \BibitemOpen
  \bibfield  {author} {\bibinfo {author} {\bibfnamefont {E.}~\bibnamefont
  {Braaten}}\ and\ \bibinfo {author} {\bibfnamefont {R.~D.}\ \bibnamefont
  {Pisarski}},\ }\bibfield  {title} {\bibinfo {title} {{Simple effective
  Lagrangian for hard thermal loops}},\ }\href
  {https://doi.org/10.1103/PhysRevD.45.R1827} {\bibfield  {journal} {\bibinfo
  {journal} {Phys. Rev. D}\ }\textbf {\bibinfo {volume} {45}},\ \bibinfo
  {pages} {R1827} (\bibinfo {year} {1992})}\BibitemShut {NoStop}%
\bibitem [{\citenamefont {Chou}\ \emph {et~al.}(1985)\citenamefont {Chou},
  \citenamefont {Su}, \citenamefont {Hao},\ and\ \citenamefont
  {Yu}}]{Chou:1984es}%
  \BibitemOpen
  \bibfield  {author} {\bibinfo {author} {\bibfnamefont {K.-c.}\ \bibnamefont
  {Chou}}, \bibinfo {author} {\bibfnamefont {Z.-b.}\ \bibnamefont {Su}},
  \bibinfo {author} {\bibfnamefont {B.-l.}\ \bibnamefont {Hao}},\ and\ \bibinfo
  {author} {\bibfnamefont {L.}~\bibnamefont {Yu}},\ }\bibfield  {title}
  {\bibinfo {title} {{Equilibrium and Nonequilibrium Formalisms Made
  Unified}},\ }\href {https://doi.org/10.1016/0370-1573(85)90136-X} {\bibfield
  {journal} {\bibinfo  {journal} {Phys. Rept.}\ }\textbf {\bibinfo {volume}
  {118}},\ \bibinfo {pages} {1} (\bibinfo {year} {1985})}\BibitemShut {NoStop}%
\bibitem [{\citenamefont {Greiner}\ and\ \citenamefont
  {Leupold}(1998)}]{Greiner:1998vd}%
  \BibitemOpen
  \bibfield  {author} {\bibinfo {author} {\bibfnamefont {C.}~\bibnamefont
  {Greiner}}\ and\ \bibinfo {author} {\bibfnamefont {S.}~\bibnamefont
  {Leupold}},\ }\bibfield  {title} {\bibinfo {title} {{Stochastic
  interpretation of Kadanoff-Baym equations and their relation to Langevin
  processes}},\ }\href {https://doi.org/10.1006/aphy.1998.5849} {\bibfield
  {journal} {\bibinfo  {journal} {Annals Phys.}\ }\textbf {\bibinfo {volume}
  {270}},\ \bibinfo {pages} {328} (\bibinfo {year} {1998})},\ \Eprint
  {https://arxiv.org/abs/hep-ph/9802312} {arXiv:hep-ph/9802312} \BibitemShut
  {NoStop}%
\bibitem [{Note1()}]{Note1}%
  \BibitemOpen
  \bibinfo {note} {$f$ is a function of $\protect \mathbf {p}$ and $\protect
  \mathbf {R}$. However, we assume that $f$ is a very smooth function in $R$
  when we look at distances of the order of $1/T$. Therefore, we can consider
  that $f$ does not depend on for the purposes of the matching and the study in
  this section. We also assume that the distribution is isotropic.}\BibitemShut
  {Stop}%
\bibitem [{\citenamefont {Baym}\ and\ \citenamefont
  {Kadanoff}(1961)}]{Baym:1961zz}%
  \BibitemOpen
  \bibfield  {author} {\bibinfo {author} {\bibfnamefont {G.}~\bibnamefont
  {Baym}}\ and\ \bibinfo {author} {\bibfnamefont {L.~P.}\ \bibnamefont
  {Kadanoff}},\ }\bibfield  {title} {\bibinfo {title} {{Conservation Laws and
  Correlation Functions}},\ }\href {https://doi.org/10.1103/PhysRev.124.287}
  {\bibfield  {journal} {\bibinfo  {journal} {Phys. Rev.}\ }\textbf {\bibinfo
  {volume} {124}},\ \bibinfo {pages} {287} (\bibinfo {year}
  {1961})}\BibitemShut {NoStop}%
\bibitem [{\citenamefont {Sheng}\ \emph {et~al.}(2021)\citenamefont {Sheng},
  \citenamefont {Weickgenannt}, \citenamefont {Speranza}, \citenamefont
  {Rischke},\ and\ \citenamefont {Wang}}]{Sheng:2021kfc}%
  \BibitemOpen
  \bibfield  {author} {\bibinfo {author} {\bibfnamefont {X.-L.}\ \bibnamefont
  {Sheng}}, \bibinfo {author} {\bibfnamefont {N.}~\bibnamefont {Weickgenannt}},
  \bibinfo {author} {\bibfnamefont {E.}~\bibnamefont {Speranza}}, \bibinfo
  {author} {\bibfnamefont {D.~H.}\ \bibnamefont {Rischke}},\ and\ \bibinfo
  {author} {\bibfnamefont {Q.}~\bibnamefont {Wang}},\ }\bibfield  {title}
  {\bibinfo {title} {{From Kadanoff-Baym to Boltzmann equations for massive
  spin-1/2 fermions}},\ }\href {https://doi.org/10.1103/PhysRevD.104.016029}
  {\bibfield  {journal} {\bibinfo  {journal} {Phys. Rev. D}\ }\textbf {\bibinfo
  {volume} {104}},\ \bibinfo {pages} {016029} (\bibinfo {year} {2021})},\
  \Eprint {https://arxiv.org/abs/2103.10636} {arXiv:2103.10636 [nucl-th]}
  \BibitemShut {NoStop}%
\bibitem [{\citenamefont {Casalderrey-Solana}\ and\ \citenamefont
  {Teaney}(2006)}]{Casalderrey-Solana:2006fio}%
  \BibitemOpen
  \bibfield  {author} {\bibinfo {author} {\bibfnamefont {J.}~\bibnamefont
  {Casalderrey-Solana}}\ and\ \bibinfo {author} {\bibfnamefont
  {D.}~\bibnamefont {Teaney}},\ }\bibfield  {title} {\bibinfo {title} {{Heavy
  quark diffusion in strongly coupled N=4 Yang-Mills}},\ }\href
  {https://doi.org/10.1103/PhysRevD.74.085012} {\bibfield  {journal} {\bibinfo
  {journal} {Phys. Rev. D}\ }\textbf {\bibinfo {volume} {74}},\ \bibinfo
  {pages} {085012} (\bibinfo {year} {2006})},\ \Eprint
  {https://arxiv.org/abs/hep-ph/0605199} {arXiv:hep-ph/0605199} \BibitemShut
  {NoStop}%
\bibitem [{\citenamefont {Ghiglieri}\ \emph {et~al.}(2020)\citenamefont
  {Ghiglieri}, \citenamefont {Kurkela}, \citenamefont {Strickland},\ and\
  \citenamefont {Vuorinen}}]{Ghiglieri:2020dpq}%
  \BibitemOpen
  \bibfield  {author} {\bibinfo {author} {\bibfnamefont {J.}~\bibnamefont
  {Ghiglieri}}, \bibinfo {author} {\bibfnamefont {A.}~\bibnamefont {Kurkela}},
  \bibinfo {author} {\bibfnamefont {M.}~\bibnamefont {Strickland}},\ and\
  \bibinfo {author} {\bibfnamefont {A.}~\bibnamefont {Vuorinen}},\ }\bibfield
  {title} {\bibinfo {title} {{Perturbative Thermal QCD: Formalism and
  Applications}},\ }\href {https://doi.org/10.1016/j.physrep.2020.07.004}
  {\bibfield  {journal} {\bibinfo  {journal} {Phys. Rept.}\ }\textbf {\bibinfo
  {volume} {880}},\ \bibinfo {pages} {1} (\bibinfo {year} {2020})},\ \Eprint
  {https://arxiv.org/abs/2002.10188} {arXiv:2002.10188 [hep-ph]} \BibitemShut
  {NoStop}%
\bibitem [{Note2()}]{Note2}%
  \BibitemOpen
  \bibinfo {note} {Propagators with a single arrow pointing to the right (left)
  are retarded (advanced). Propagators with two outgoing arrows and a \protect
  \textit {capacitor} symbol are symmetric. The rest of the symbols are defined
  in appendix \ref {sec:frnrqcdl}}\BibitemShut {NoStop}%
\bibitem [{\citenamefont {Caron-Huot}\ \emph {et~al.}(2009)\citenamefont
  {Caron-Huot}, \citenamefont {Laine},\ and\ \citenamefont
  {Moore}}]{Caron-Huot:2009ncn}%
  \BibitemOpen
  \bibfield  {author} {\bibinfo {author} {\bibfnamefont {S.}~\bibnamefont
  {Caron-Huot}}, \bibinfo {author} {\bibfnamefont {M.}~\bibnamefont {Laine}},\
  and\ \bibinfo {author} {\bibfnamefont {G.~D.}\ \bibnamefont {Moore}},\
  }\bibfield  {title} {\bibinfo {title} {{A Way to estimate the heavy quark
  thermalization rate from the lattice}},\ }\href
  {https://doi.org/10.1088/1126-6708/2009/04/053} {\bibfield  {journal}
  {\bibinfo  {journal} {JHEP}\ }\textbf {\bibinfo {volume} {04}},\ \bibinfo
  {pages} {053}},\ \Eprint {https://arxiv.org/abs/0901.1195} {arXiv:0901.1195
  [hep-lat]} \BibitemShut {NoStop}%
\bibitem [{\citenamefont {Blaizot}\ and\ \citenamefont
  {Escobedo}(2018)}]{Blaizot:2017ypk}%
  \BibitemOpen
  \bibfield  {author} {\bibinfo {author} {\bibfnamefont {J.-P.}\ \bibnamefont
  {Blaizot}}\ and\ \bibinfo {author} {\bibfnamefont {M.~A.}\ \bibnamefont
  {Escobedo}},\ }\bibfield  {title} {\bibinfo {title} {{Quantum and classical
  dynamics of heavy quarks in a quark-gluon plasma}},\ }\href
  {https://doi.org/10.1007/JHEP06(2018)034} {\bibfield  {journal} {\bibinfo
  {journal} {JHEP}\ }\textbf {\bibinfo {volume} {06}},\ \bibinfo {pages}
  {034}},\ \Eprint {https://arxiv.org/abs/1711.10812} {arXiv:1711.10812
  [hep-ph]} \BibitemShut {NoStop}%
\bibitem [{\citenamefont {Delorme}(2021)}]{Delorme:2021eno}%
  \BibitemOpen
  \bibfield  {author} {\bibinfo {author} {\bibfnamefont {S.}~\bibnamefont
  {Delorme}},\ }\emph {\bibinfo {title} {{Theoretical description of quarkonium
  dynamics in the quark gluon plasma with a quantum master equation
  approach}}},\ \href@noop {} {Ph.D. thesis},\ \bibinfo  {school} {Laboratoire
  de physique subatomique et des technologies associ\'ees, France, IMT
  Atlantique} (\bibinfo {year} {2021})\BibitemShut {NoStop}%
\bibitem [{\citenamefont {Brambilla}\ \emph {et~al.}(2022)\citenamefont
  {Brambilla}, \citenamefont {Escobedo}, \citenamefont {Islam}, \citenamefont
  {Strickland}, \citenamefont {Tiwari}, \citenamefont {Vairo},\ and\
  \citenamefont {Vander~Griend}}]{Brambilla:2022ynh}%
  \BibitemOpen
  \bibfield  {author} {\bibinfo {author} {\bibfnamefont {N.}~\bibnamefont
  {Brambilla}}, \bibinfo {author} {\bibfnamefont {M.~A.}\ \bibnamefont
  {Escobedo}}, \bibinfo {author} {\bibfnamefont {A.}~\bibnamefont {Islam}},
  \bibinfo {author} {\bibfnamefont {M.}~\bibnamefont {Strickland}}, \bibinfo
  {author} {\bibfnamefont {A.}~\bibnamefont {Tiwari}}, \bibinfo {author}
  {\bibfnamefont {A.}~\bibnamefont {Vairo}},\ and\ \bibinfo {author}
  {\bibfnamefont {P.}~\bibnamefont {Vander~Griend}},\ }\bibfield  {title}
  {\bibinfo {title} {{Heavy quarkonium dynamics at next-to-leading order in the
  binding energy over temperature}},\ }\href
  {https://doi.org/10.1007/JHEP08(2022)303} {\bibfield  {journal} {\bibinfo
  {journal} {JHEP}\ }\textbf {\bibinfo {volume} {08}},\ \bibinfo {pages}
  {303}},\ \Eprint {https://arxiv.org/abs/2205.10289} {arXiv:2205.10289
  [hep-ph]} \BibitemShut {NoStop}%
\bibitem [{\citenamefont {Brambilla}\ \emph {et~al.}(2023)\citenamefont
  {Brambilla}, \citenamefont {Escobedo}, \citenamefont {Islam}, \citenamefont
  {Strickland}, \citenamefont {Tiwari}, \citenamefont {Vairo},\ and\
  \citenamefont {Vander~Griend}}]{Brambilla:2023hkw}%
  \BibitemOpen
  \bibfield  {author} {\bibinfo {author} {\bibfnamefont {N.}~\bibnamefont
  {Brambilla}}, \bibinfo {author} {\bibfnamefont {M.~A.}\ \bibnamefont
  {Escobedo}}, \bibinfo {author} {\bibfnamefont {A.}~\bibnamefont {Islam}},
  \bibinfo {author} {\bibfnamefont {M.}~\bibnamefont {Strickland}}, \bibinfo
  {author} {\bibfnamefont {A.}~\bibnamefont {Tiwari}}, \bibinfo {author}
  {\bibfnamefont {A.}~\bibnamefont {Vairo}},\ and\ \bibinfo {author}
  {\bibfnamefont {P.}~\bibnamefont {Vander~Griend}},\ }\bibfield  {title}
  {\bibinfo {title} {{Regeneration of bottomonia in an open quantum systems
  approach}},\ }\href {https://doi.org/10.1103/PhysRevD.108.L011502} {\bibfield
   {journal} {\bibinfo  {journal} {Phys. Rev. D}\ }\textbf {\bibinfo {volume}
  {108}},\ \bibinfo {pages} {L011502} (\bibinfo {year} {2023})},\ \Eprint
  {https://arxiv.org/abs/2302.11826} {arXiv:2302.11826 [hep-ph]} \BibitemShut
  {NoStop}%
\bibitem [{\citenamefont {Veltman}(1963)}]{Veltman:1963th}%
  \BibitemOpen
  \bibfield  {author} {\bibinfo {author} {\bibfnamefont {M.~J.~G.}\
  \bibnamefont {Veltman}},\ }\bibfield  {title} {\bibinfo {title} {{Unitarity
  and causality in a renormalizable field theory with unstable particles}},\
  }\href {https://doi.org/10.1016/S0031-8914(63)80277-3} {\bibfield  {journal}
  {\bibinfo  {journal} {Physica}\ }\textbf {\bibinfo {volume} {29}},\ \bibinfo
  {pages} {186} (\bibinfo {year} {1963})}\BibitemShut {NoStop}%
\bibitem [{\citenamefont {Ruegg}\ and\ \citenamefont
  {Ruiz-Altaba}(2004)}]{Ruegg:2003ps}%
  \BibitemOpen
  \bibfield  {author} {\bibinfo {author} {\bibfnamefont {H.}~\bibnamefont
  {Ruegg}}\ and\ \bibinfo {author} {\bibfnamefont {M.}~\bibnamefont
  {Ruiz-Altaba}},\ }\bibfield  {title} {\bibinfo {title} {{The Stueckelberg
  field}},\ }\href {https://doi.org/10.1142/S0217751X04019755} {\bibfield
  {journal} {\bibinfo  {journal} {Int. J. Mod. Phys. A}\ }\textbf {\bibinfo
  {volume} {19}},\ \bibinfo {pages} {3265} (\bibinfo {year} {2004})},\ \Eprint
  {https://arxiv.org/abs/hep-th/0304245} {arXiv:hep-th/0304245} \BibitemShut
  {NoStop}%
\end{thebibliography}%
\end{document}